\documentclass[preprint]{aastex}
\usepackage{epstopdf}
\usepackage{natbib}
\usepackage{amsmath}
\usepackage{graphicx}

\begin{document}
\title{Panchromatic Hubble Andromeda Treasury IX: A Photometric Survey of Planetary Nebulae
in M31}
\shorttitle{PHAT IX: A Photometric Survey of PNe in M31}

\author{Mark J. Veyette\altaffilmark{1,2},
Benjamin F. Williams\altaffilmark{1},
Julianne J. Dalcanton\altaffilmark{1},
Bruce Balick\altaffilmark{1},
Nelson Caldwell\altaffilmark{3},
Morgan Fouesneau\altaffilmark{1,4},
L\'eo Girardi\altaffilmark{5},
Karl D. Gordon\altaffilmark{6,7},
Jason Kalirai\altaffilmark{6,8},
Philip Rosenfield\altaffilmark{9},
Anil C. Seth\altaffilmark{10}
}
\altaffiltext{1}{Department of Astronomy, University of Washington, Box 351580, Seattle, WA
98195, USA}
\altaffiltext{2}{Astronomy Department, Boston University, 725 Commonwealth Ave, Boston, MA,
02215, USA}
\altaffiltext{3}{Harvard-Smithsonian Center for Astrophysics, 60 Garden Street Cambridge, MA
02138, USA}
\altaffiltext{4}{Max-Planck-Institut fur Astronomie, K{\"o}nigstuhl 17, 69117 Heidelberg, Germany}
\altaffiltext{5}{Osservatorio Astronomico di Padova -- INAF, Vicolo dell'Osservatorio 5,
I-35122 Padova, Italy}
\altaffiltext{6}{Space Telescope Science Institute, 3700 San Martin Drive, Baltimore, MD,
21218, USA}
\altaffiltext{7}{Sterrenkundig Observatorium, Universiteit Gent, Gent, Belgium}
\altaffiltext{8}{Center for Astrophysical Sciences, Johns Hopkins University, Baltimore MD
21218}
\altaffiltext{9}{Department of Physics and Astronomy G. Galilei, University of Padova, Vicolo
dell'Osservatorio 3, I-35122 Padova, Italy}
\altaffiltext{10}{Department of Physics \& Astronomy, University of Utah, Salt Lake City, UT
84112}

\shortauthors{Veyette, Williams, Dalcanton, et al.}


\begin{abstract}

We search Hubble Space Telescope (HST) Advanced Camera for Surveys
(ACS) and Wide Field Camera 3 (WFC3) broadband imaging data from the
Panchromatic Hubble Andromeda Treasury (PHAT) survey to identify
detections of cataloged planetary nebulae (PNe). Of the 711 PNe
currently in the literature within the PHAT footprint, we find 467
detected in the broadband. For these 467 we are able to refine their
astrometric accuracy from $\sim$0$\farcs3$ to $0\farcs05$. Using the resolution of HST,
we are able to show that 152 objects currently in the catalogs are
definitively not PNe, and we show that 32 objects thought to be
extended in ground-based images are actually point-like and therefore
good PN candidates. We also find one PN candidate that is marginally
resolved. If this is a PN, it is up to 0.7 pc in diameter. With our new photometric data, we
develop a method of measuring the level of excitation in individual PNe by comparing broadband
and narrowband imaging and describe the effects of excitation on a PN's photometric signature.
Using the photometric properties of the
known PNe in the PHAT catalogs, we search for more PN, but do not find
any new candidates, suggesting that ground-based emission-line surveys
are complete in the PHAT footprint to $F475W\simeq24$.

\end{abstract}

\section{Introduction}\label{introduction}

As the nearest large spiral galaxy, M31 offers a unique opportunity to study planetary nebulae
(PNe) on a galactic scale. Stars born with masses between 0.8-8 $M_{\sun}$ will likely go
through a planetary nebula phase, making PNe good tracers of old stellar populations. Due to
the short time scales between the AGB and PN phases, the distribution and kinematics of PNe
are expected to be identical to their parent population. Studying PNe as a population provides
insight into galactic structure and evolution. Understanding their evolution individually can
provide insight into PN structure, formation, and enrichment of the interstellar medium and
provide constraints for future PN evolutionary models. Studying PNe has the additional benefit
of ease of detection and classification due to their strong emission features and unique
spectra. Further, the population of PNe in M31 all lie at the same distance, making it
possible to directly relate the luminosity and emission of a large and diverse sample of PNe.

Previous surveys have identified and studied thousands of PNe throughout the disk and bulge of
M31 \citep{ciardullo1989,hurley2004,halliday2006,merrett2006}. Recent spectroscopic studies
have begun to dig deeper into their physical properties \citep{kwitter2012,sanders2012}. The
largest survey of M31 PNe to date is a deep kinematic survey performed by \cite{merrett2006}
(hereafter M06) using the purpose-built Planetary Nebula Spectrograph \citep{douglas2002},
which can simultaneously detect PNe and measure their radial velocities. M06 identified 3300
emission-line objects in M31, of which 2615 are likely to be PNe.

Single-purpose spectroscopic surveys are very useful for studying the properties of individual
PNe, but can be time consuming and costly. In contrast, multi-purpose photometric surveys can
offer substantial, albeit less detailed, information about previously discovered PNe without
the need for additional observations and can even be used to identify previously undiscovered
PNe \citep{kniazev2014}. Broad spectral coverage allows one to probe large spans of the
spectral energy distributions of PNe and can be used in concert with spectroscopic surveys to 
calibrate photometric data for the study of PN emission. Although this photometric information
alone is not as detailed as spectroscopic data, it is sufficient to place broad constraints on
the properties of the overall strength of the PN's emission lines and underlying stellar
continuum for much larger populations than are accessible spectroscopically.

PNe are typically identified by searching for objects with bright [\ion{O}{3}] $\lambda$5007
lines. Traditional large area surveys use the difference of on- and off-band narrowband
photometry to select PN candidates which are then followed up spectroscopically to exclude
\ion{H}{2} regions and Wolf-Rayet stars, which also emit strong [\ion{O}{3}] lines. Ground
based spectroscopy is limited to only the brightest PNe and restricted by atmospheric windows.
For the M06 survey, spectroscopic follow up was replaced by cuts of extended objects and
objects with low $F_{5007}/F_{H\alpha}$ ratios measured from narrowband imaging from the Local
Group Survey (LGS) \citep{massey2006}. While this method provided a deeper and larger survey
without the need for spectroscopic follow-up, it is also subject to more contamination. High
resolution imaging provides a cleaner view of the sources of the [\ion{O}{3}] emission,
improving rejection of non-PN sources, particularly extended \ion{H}{2} regions which are the
dominant contaminating object in PN surveys.

In Section~\ref{data} of this paper we describe how we identified PNe in the PHAT catalog by
cross-identification with M06. In Section~\ref{results} we present the PHAT PNe Catalog and
characterize basic UV, optical, and NIR photometric properties of PNe. We discuss our results
in Section~\ref{discussion} and summarize them in Section~\ref{summary}.

\section{Data Acquisition and Analysis}\label{data}

\subsection{The Panchromatic Hubble Andromeda Treasury}
The Panchromatic Hubble Andromeda Treasury (PHAT) is a Hubble Space Telescope (HST) multi
cycle imaging survey covering roughly 1/3 of M31's star forming disk out to 20 kpc
\citep{dalcanton2012}. Observations were made with HST using the Advanced Camera for Surveys
(ACS) and Wide Field Camera 3 (WFC3) with six filters spanning the ultraviolet (WFC3/UVIS
F275W and F336W), the optical (ACS/WFC F475W and F814W), and the near-infrared (WFC3/IR F110W
and F160W). All data files are publicly available and described at
\url{http://archive.stsci.edu/prepds/phat/}. For full details of the data acquisition and
analysis, see \citet{dalcanton2012}.

Briefly, photometry was reduced using the DOLPHOT 1.2 \citep{dolphin2000} software package
which uses iterative point spread function (PSF)-fitting photometry on neighbor-subtracted
images to fit the sky and PSF simultaneously. This process is done for each peak in a stack of
images across different filters of the same camera. DOLPHOT aligns each image in a stack using
hundreds to thousands of bright stars in the field to achieve the precise alignment needed for
its photometric processing. Further internal astrometric alignment was performed for each
camera. The cameras were then aligned with each other and with a global reference frame. The
reference frame was provided by archival $i^{\prime}$ data from MegaCam CFHT. The error in
ACS/WFC to CFHT alignment is $0\farcs05$ and the relative error in alignment across the three
cameras is $0\farcs01$. Cross identification with M06 is therefore limited by the astrometric
accuracy of M06 --- reported to be $0\farcs34$ and $0\farcs16$ in RA and Dec respectively.

We avoided restricting our search to the PHAT point source catalogs. Rather, we cross
identified PNe using the ACS ``{\tt{*.phot}}'' files produced by DOLPHOT. We used these raw
DOLPHOT output files which contain sources that may poorly fit the PSF so as not to exclude
any extended PNe.

PHAT is a multi-release program that is still in progress. This paper covers only the PHAT
data available at the time the work for this paper was done. PHAT consists of a total of 23
bricks, each containing 18 HST pointings. This paper covers only Bricks 2, 4, 5, 6, 8, 9, 12,
14, 15, 16, 17, 18, 19, 21, 22, and 23. Brick 1 was excluded from this paper, though currently
available, due to heavy crowding effects that made it difficult to identify PNe in the inner
regions of M31.

\subsection{Identifying PNe in PHAT}
All PNe analyzed in this paper were originally identified as part of the M06 deep kinematic
survey of M31, which covered M31's disk out to a radius of 2$^{\circ}$ (27.4 kpc). M06
identified 2615 likely PNe down to a reported completeness of $m5007\simeq23.75$ for the
[\ion{O}{3}] $\lambda$5007 line. Of these 2615, 711 are in the current PHAT footprint.
Individual PN candidates have reported uncertainties in their $m5007$ measurements of 0.07 mag
and in their RA and Dec measurements of $0\farcs34$ and $0\farcs16$ respectively. We cross
identified sources in the M06 catalog with photometric data from PHAT.

\subsubsection{PNe in the PHAT filters}
\begin{figure}
\centering
\includegraphics[width=7in]{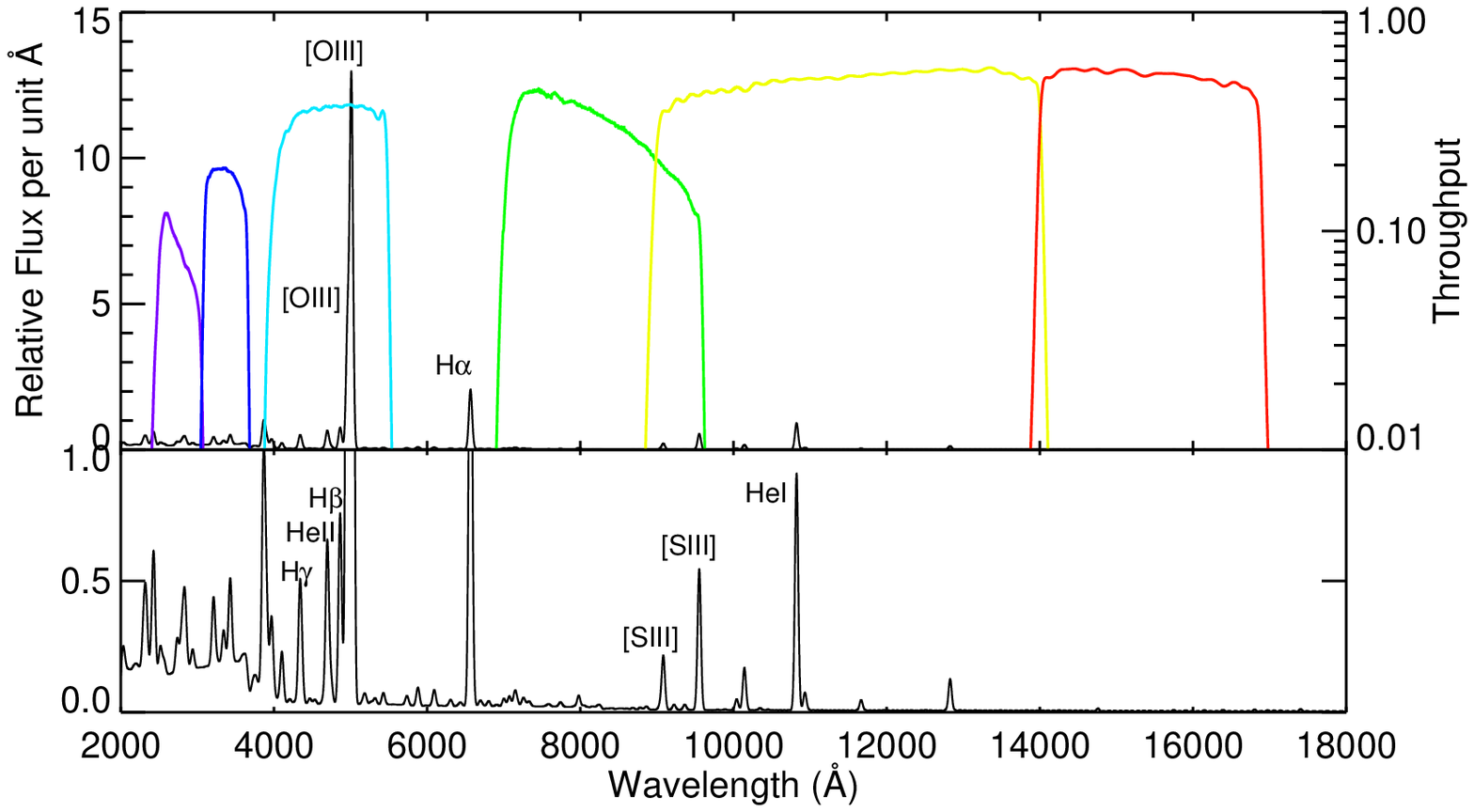}
\caption{(Top) A model spectrum (black line) for a PN model close to its maximum [\ion{O}{3}]
$\lambda$5007 luminosity and with high excitation. This particular model was extracted from
the \cite{marigo2001,marigo2004} database of simplified evolutionary models, in which the
photoionization of a spherically symmetric expanding shell is simulated with
\cite{ferland2013} Cloudy code v08.01. Some of the main contributors to the nebular flux are
marked in the figure. In addition, there is a significant flux contribution from the stellar
and nebular continuum, especially in the UV, as listed in Table~\ref{fluxtable}. Also shown
are filter throughput curves (colored lines) for filters: (from left to right) F275W, F336W,
F475W, F814W, F110W, and F160W. (Bottom) A closer look at the continuum and weaker emission
lines.}
\label{throughput}
\end{figure}

Figure~\ref{throughput} shows throughput curves for each of the PHAT filters overlaid on a
model PN spectrum. The bandpasses of the two UV filters, F275W and F336W, overlie many weak
emission lines together with a significant stellar and nebular continuum \citep[see][]
{bianchi1997}. The majority of nebular emission line flux falls under the bandpass of the
F475W filter, namely the strong emission lines: [\ion{O}{3}] $\lambda$5007, [\ion{O}{3}]
$\lambda$4959, $H\beta$, and $H\gamma$. Therefore, F475W flux is due largely to nebular
emission lines. The F814W bandpass overlies a few weak emission lines, though the majority of
flux in F814W is likely due to stellar and Paschen continuum. A few weak emission lines also
fall under the F110W bandpass, which could have a significant or dominant contribution to the
total flux given the weaker underlying stellar continuum in the NIR. Flux in F160W, however,
is most likely entirely due to continuum from the central star as there are no emission lines
that could have a significant contribution to the total F160W flux. Table \ref{fluxtable}
lists likely main contributors of PNe flux in each of the PHAT filters.

\begin{figure}
	\begin{minipage}[t]{.3\textwidth}
		\centering
		\includegraphics[width=\textwidth]{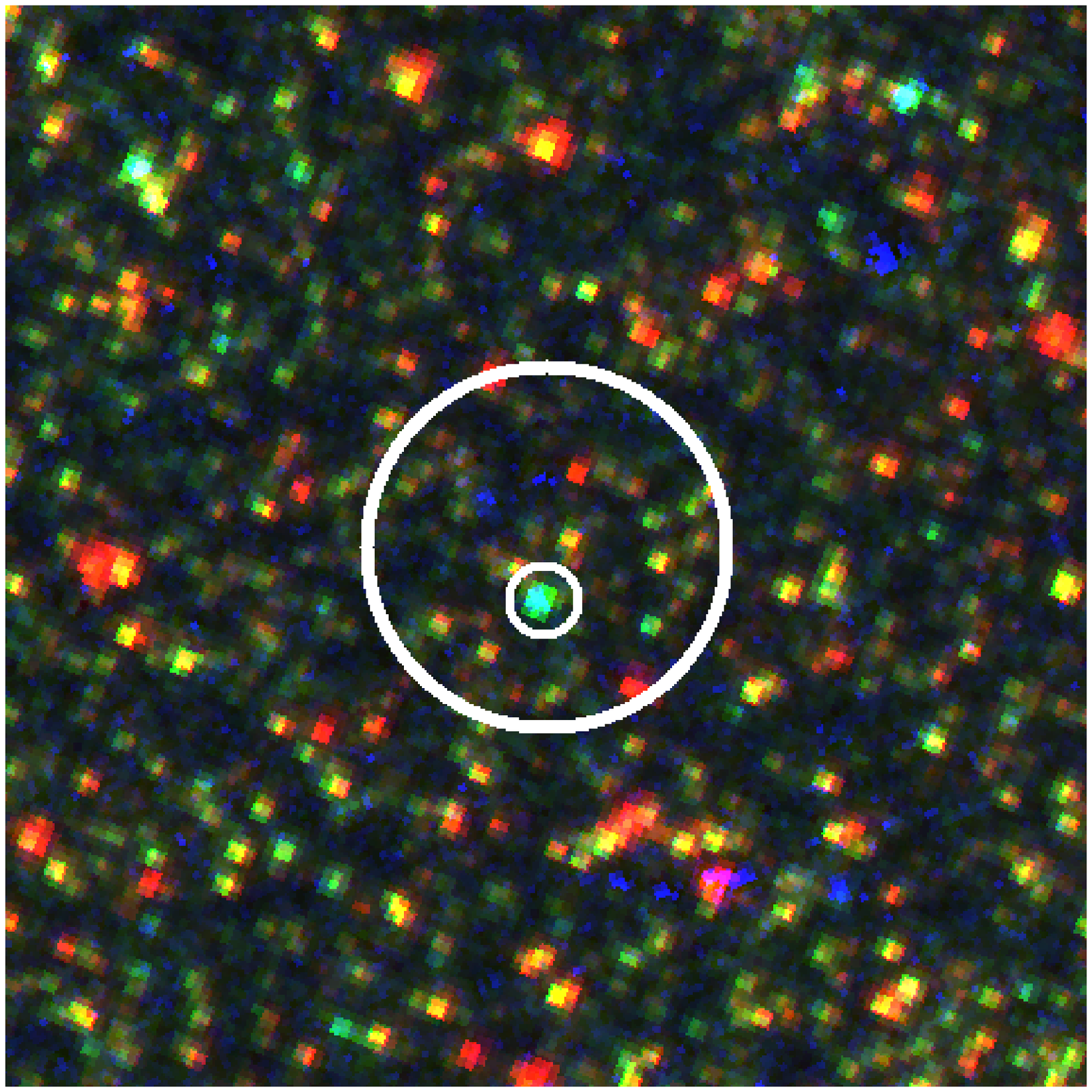}
		\includegraphics[width=\textwidth]{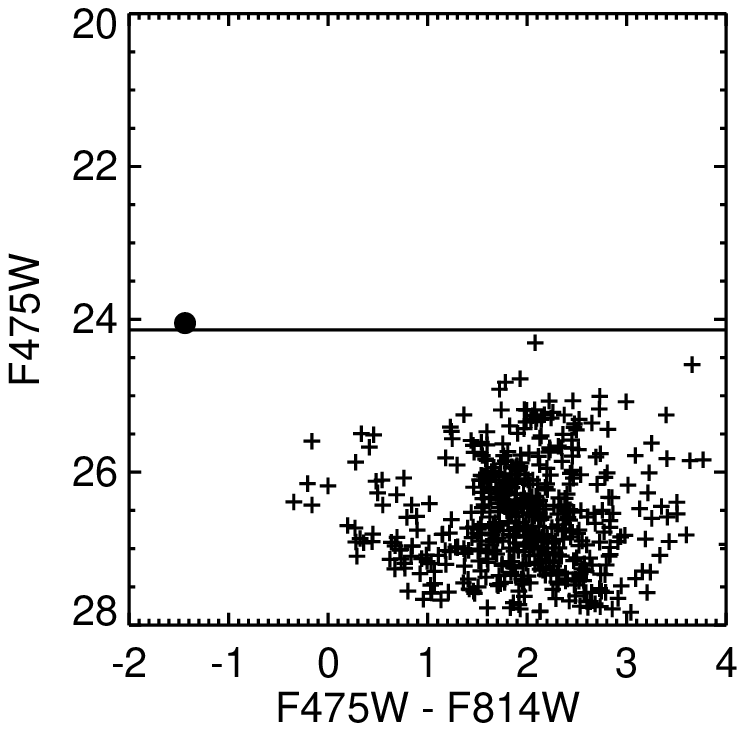}
	\end{minipage}
	\begin{minipage}[t]{.3\textwidth}
		\centering
		\includegraphics[width=\textwidth]{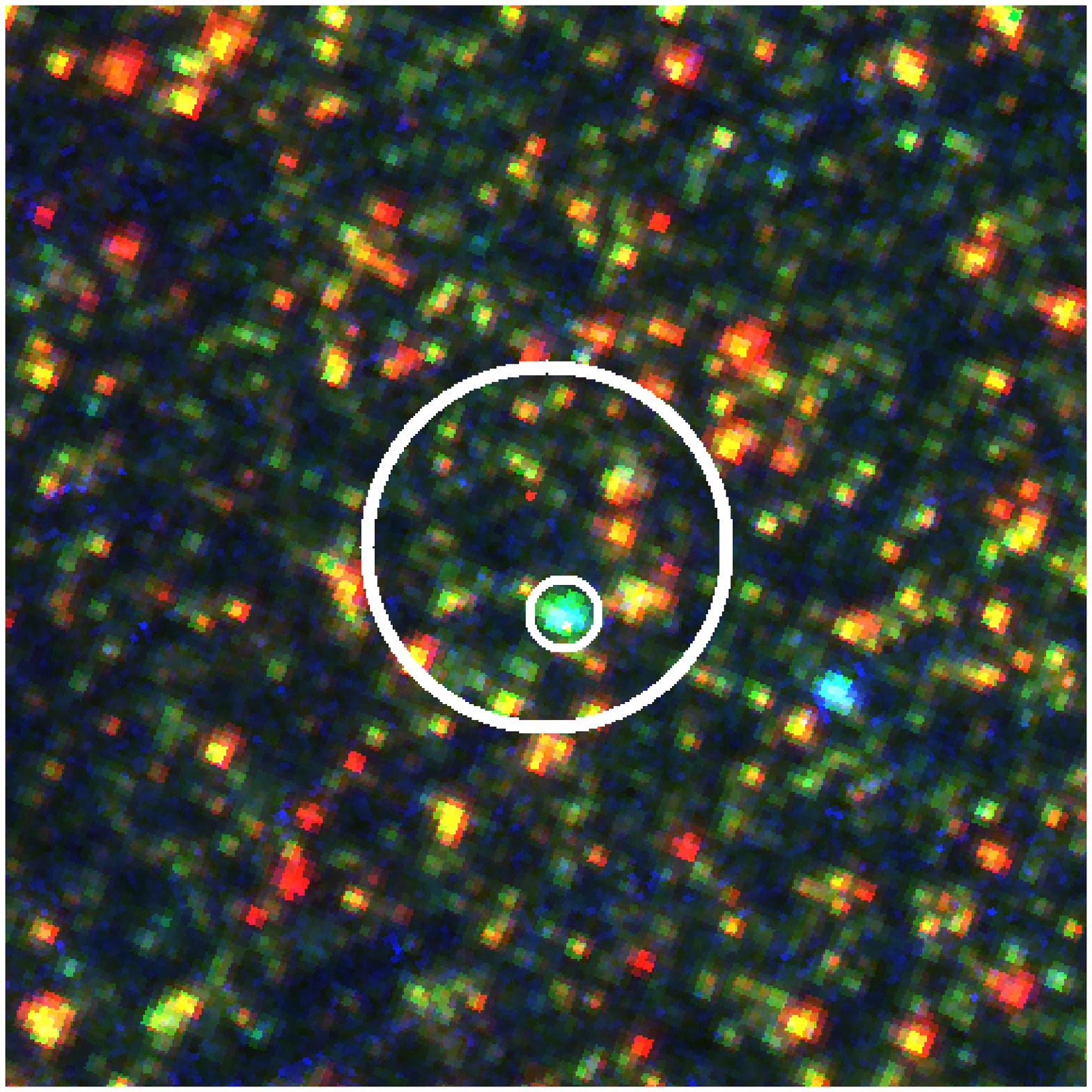}
		\includegraphics[width=\textwidth]{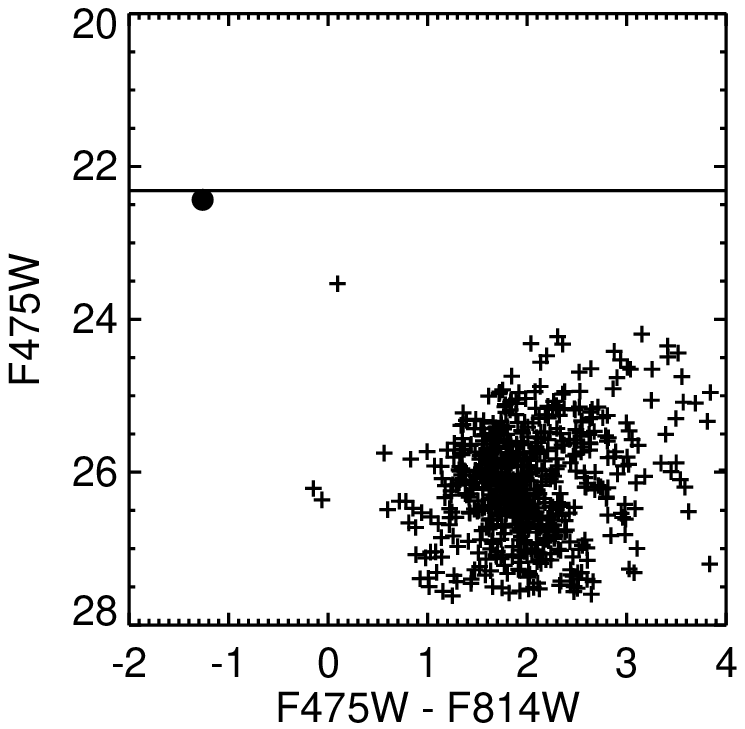}
	\end{minipage}
	\begin{minipage}[t]{.3\textwidth}
		\centering
		\includegraphics[width=\textwidth]{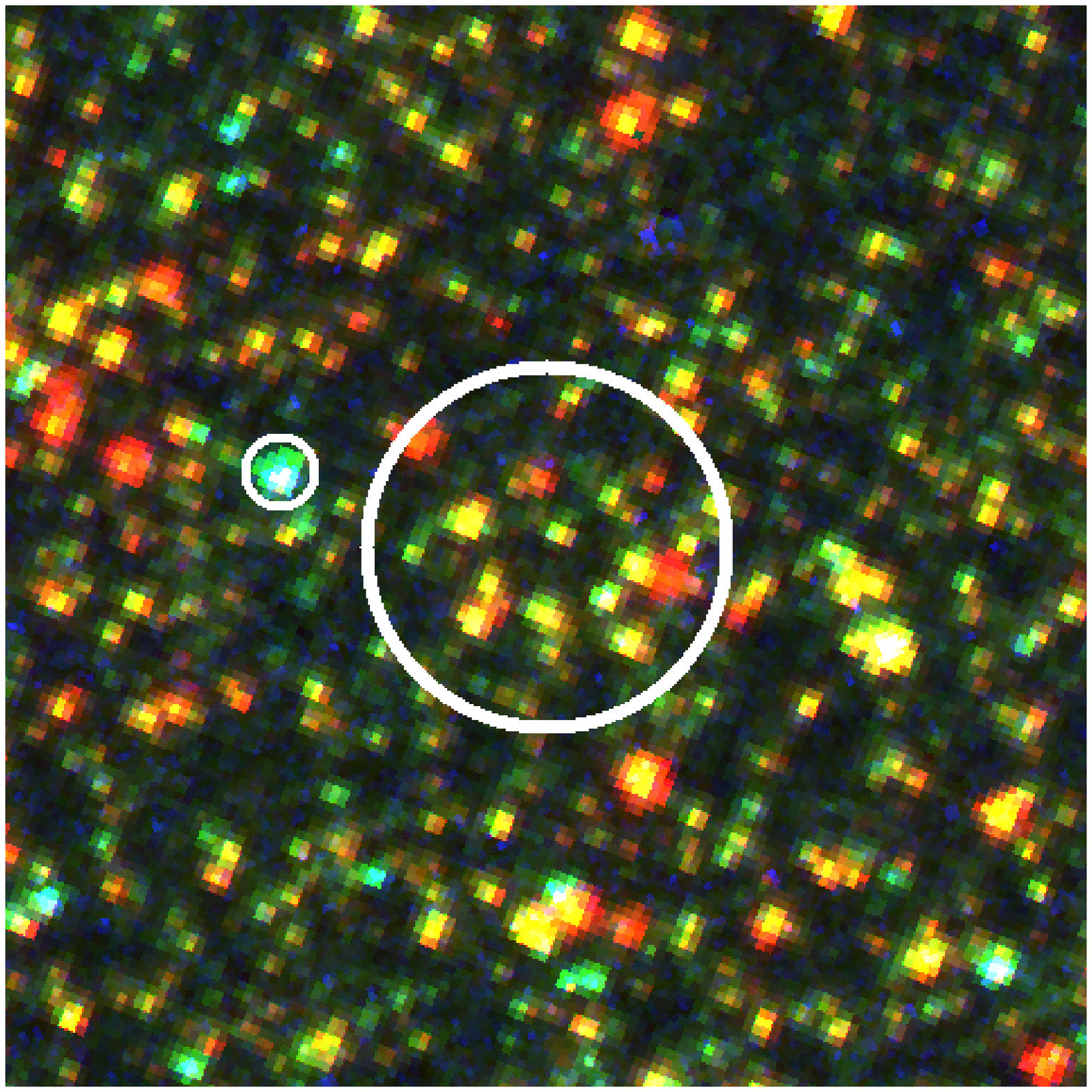}
		\includegraphics[width=\textwidth]{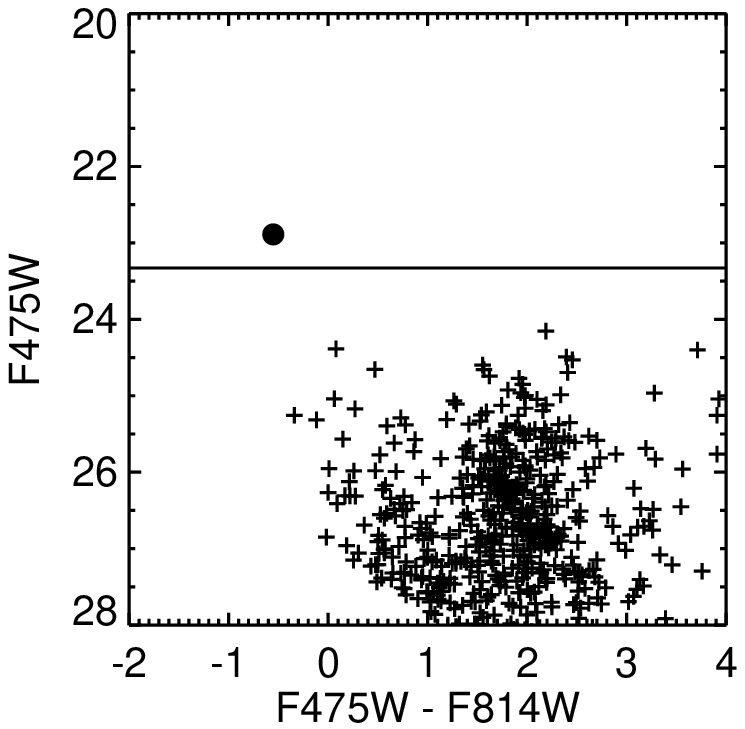}
	\end{minipage}
	\caption{(Above) 3-color images of 3 PNe in PHAT. F814W, F475W, and F336W were used for
	the red, green, and blue images respectively so that PNe appear as bright, blue-green
	objects. PNe are marked by a small white circle. A larger white circle with a radius of
	$1\arcsec$ is centered at the M06 RA and Dec for each PN. (Below) The respective optical
	CMD with a filled, black circle denoting the PN and pluses denoting stars within
	$3\arcsec$ of the PN. A black line denotes the expected F475W magnitude estimated from the
	M06 $m5007$. From left to right the sources are M06 3015 (a typical PN), M06 938 (a
	bright, crowded PN), and M06 2665 (a bright, offset PN).}
	\label{pnpostage}
\end{figure}

The vast majority of PN emission line flux within the PHAT spectral range falls in the F475W
bandpass. Thus, most PNe have unusual colors, particularly in three-color images where F475W
drives the green channel. Figure~\ref{pnpostage} shows F814W:F475W:F336W R:G:B images in which
PNe appear anomalously blue-green, making them easy to visually differentiate from surrounding
stars. We generated equivalent images for all the PHAT data and used them to visually identify
PNe, as described in the following section.

\subsubsection{PHAT Astrometry and Photometry}\label{zPNsect}
To assess the relative positional accuracy of the M06 and PHAT catalogs, we performed an
initial visual search for anomalously blue-green objects within 3 times the $1\sigma$
uncertainties of the M06 reported positions. We found likely optical counterparts as far as
$2\arcsec$ from their cataloged position. We therefore adopted $3\arcsec$ (11 pc at the distance of M31) as a search radius.

\begin{figure}
\centering
\includegraphics[width=4in]{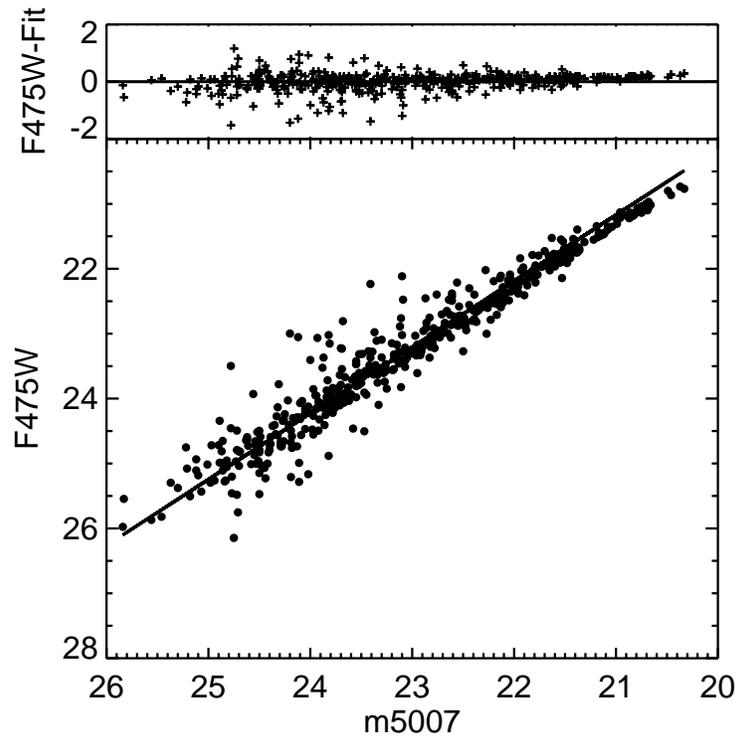}
\caption{The linear relation between $m5007$ and F475W magnitudes of the final PHAT PNe
catalog. This plot shows the strong dependence of F475W magnitude on [\ion{O}{3}]
$\lambda$5007 line strength. The linear fit was calculated taking into account the quoted M06
$m5007$ uncertainty as well as the PHAT Poisson errors from F475W photometry.}
\label{m5007vf475w}
\end{figure}

There are, on average, $\sim$600 objects in PHAT within $3\arcsec$ of an M06 PN location. We,
therefore, require additional constraints to identify the PHAT counterpart. We first take
advantage of the fact that there is a strong linear relation between the quoted M06 $m5007$
magnitude and the PHAT F475W magnitude, since the majority of F475W flux is due to the
[\ion{O}{3}] $\lambda$5007 line. We show this relation in Figure~\ref{m5007vf475w} for the
final catalog. A linear fit to this relation finds $F475W = -0.2240 + 1.0187 \times m5007$. In
addition to magnitude, we also considered cross-identifying sources based on optical color,
positional offset from M06, and the ``{\tt{sharpness}}'' and ``{\tt{roundness}}'' of the PSF.
The most distinguishing identification parameters were optical color and the expected F475W
magnitude estimated from the M06 $m5007$ magnitude. The round and sharp PSF values from PHAT's
photometry were used mainly to weed out artifacts and spurious objects such as cosmic rays
from the uncut ``{\tt{*.phot}}'' files.

To automatically select candidates on the basis of the above parameters, we used an initial
training set of visually identified PNe. We calculated the average value and $1\sigma$ spread
of each identification parameter (expected F475W, optical color, positional offset from M06,
sharpness, and roundness). For each object in PHAT within $3\arcsec$ of an M06 PN location,
the differences between the training set's and the object's parameter values were normalized
by the $1\sigma$ spreads of the parameter values and then averaged over all parameters giving
double the weight to the optical color and estimated F475W magnitude. The resulting value,
noted as $z_{PN}$, is the object's standard score of the deviation from typical PNe 
identified in PHAT. Objects with $z_{PN} < 1$ are highly probable PN candidates. Objects with 
$z_{PN} < 1.5$ are probable PN candidates. Objects with $z_{PN} < 3$ are possible PN 
candidates. Eqs.~\ref{zscore} and \ref{zpn} define $z_{PN}$ where, for parameter $x$, $x_{obj}$ 
is the object's value, $\bar{x}_{PN}$ is the training set's average value, and 
$\sigma_{x_{PN}}$ is the $1\sigma$ spread.

\begin{align}
z_{x} \equiv & \frac{|x_{obj} - \bar{x}_{PN}|}{\sigma_{x_{PN}}} \label{zscore} \\
z_{PN} \equiv & (2 z_{F475W} + 2 z_{color} + z_{RA} \nonumber \\
&+ z_{Dec} + z_{sharp} + z_{round}) / 8 \label{zpn}
\end{align}

We attempted to identify new PNe in M31 using the optical color and F475W magnitude space
occupied by the M06 PNe. We looked for [\ion{O}{3}] emission in the LGS [\ion{O}{3}] images
\citep{massey2006} to confirm these candidate PNe. None of these candidates had measurable
[\ion{O}{3}] emission. Therefore, this attempt did not result in the identification of any new
PNe, suggesting M06 is complete to the LGS [\ion{O}{3}] detection limit. An earlier attempt to
photometrically identify new PNe in PHAT was made by selecting sources in the bulge of M31
with $F475W<24.0$, $F475W-F814W<-0.3$, and no matching source in the literature. Spectroscopic
follow-up of 16 of these candidates found that 7 showed emission lines indicative of PNe. A
larger campaign could lead to the discovery of PNe in the inner regions of M31, where previous
surveys suffered extreme crowding.

Finally, all catalogs and figures presented contain only photometric measurements with a
signal-to-noise ratio (SNR) greater than 4. This SNR cut, which is necessary to ensure only
accurate, high-quality measurements are analyzed, resulted in the culling of 120 F275W
measurements, 2 F336W measurements, 1 F814W measurement, and 29 F160W measurements.

\subsubsection{Visual Assessment}
To aid in the assessment of our automatic selection method, a visual catalog was created
containing a PHAT 3-color image, PHAT optical CMD, LGS [\ion{O}{3}] image, and LGS V-band
image \citep{massey2006}, for each M06 PN location that PHAT overlaps. The three objects with
the lowest $z_{PN}$ were marked in the PHAT image and CMD. A final PNe catalog was
created using the one probable candidate in fields where there was only one probable
candidate, and the bluest of the candidates with the three lowest $z_{PN}$ in the cases
where there were multiple probable candidates or no probable candidates. This final catalog of
467 PNe is presented in Section~\ref{results}.

\begin{figure}
	\begin{minipage}[t]{.3\textwidth}
		\centering
		\includegraphics[width=\textwidth]{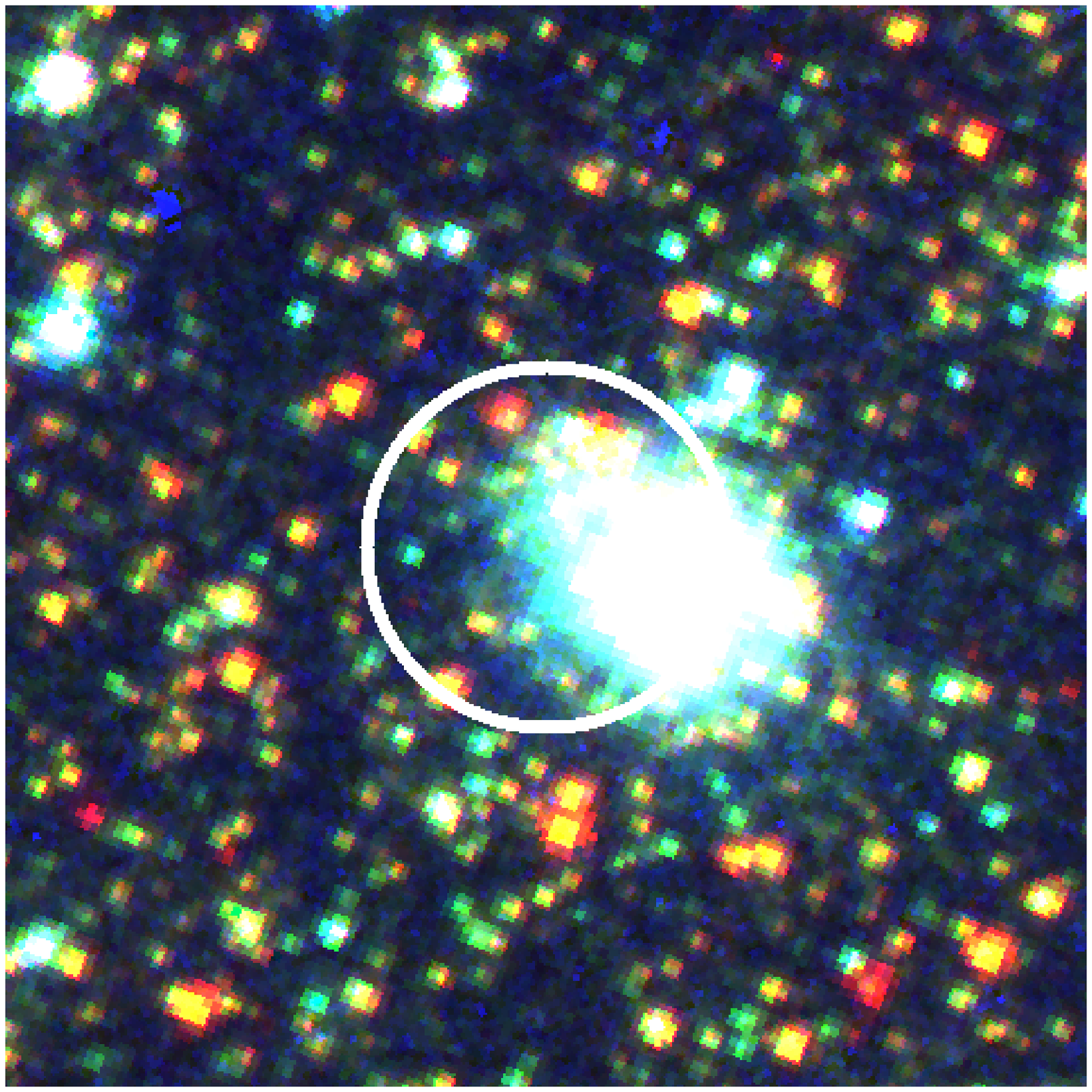}
		\includegraphics[width=\textwidth]{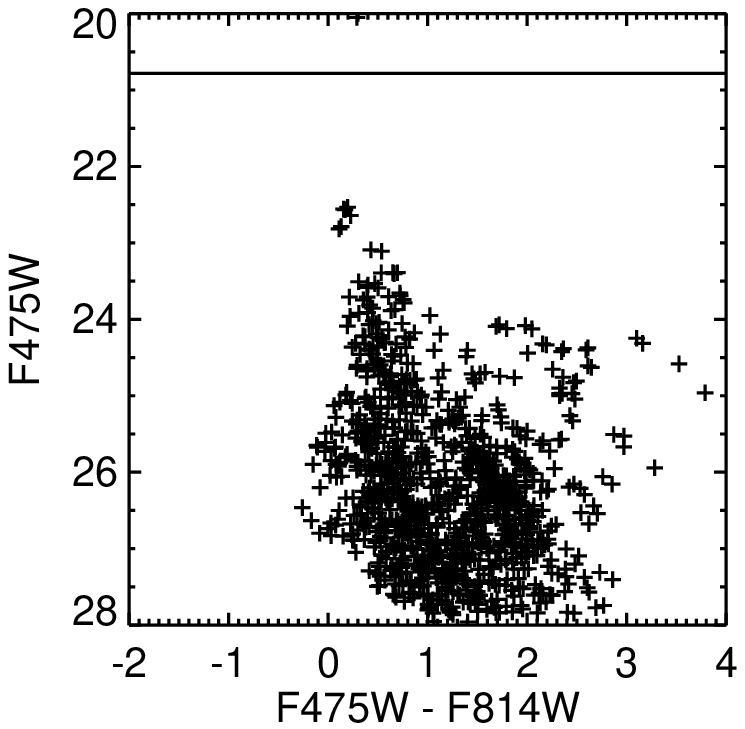}
	\end{minipage}
	\begin{minipage}[t]{.3\textwidth}
		\centering
		\includegraphics[width=\textwidth]{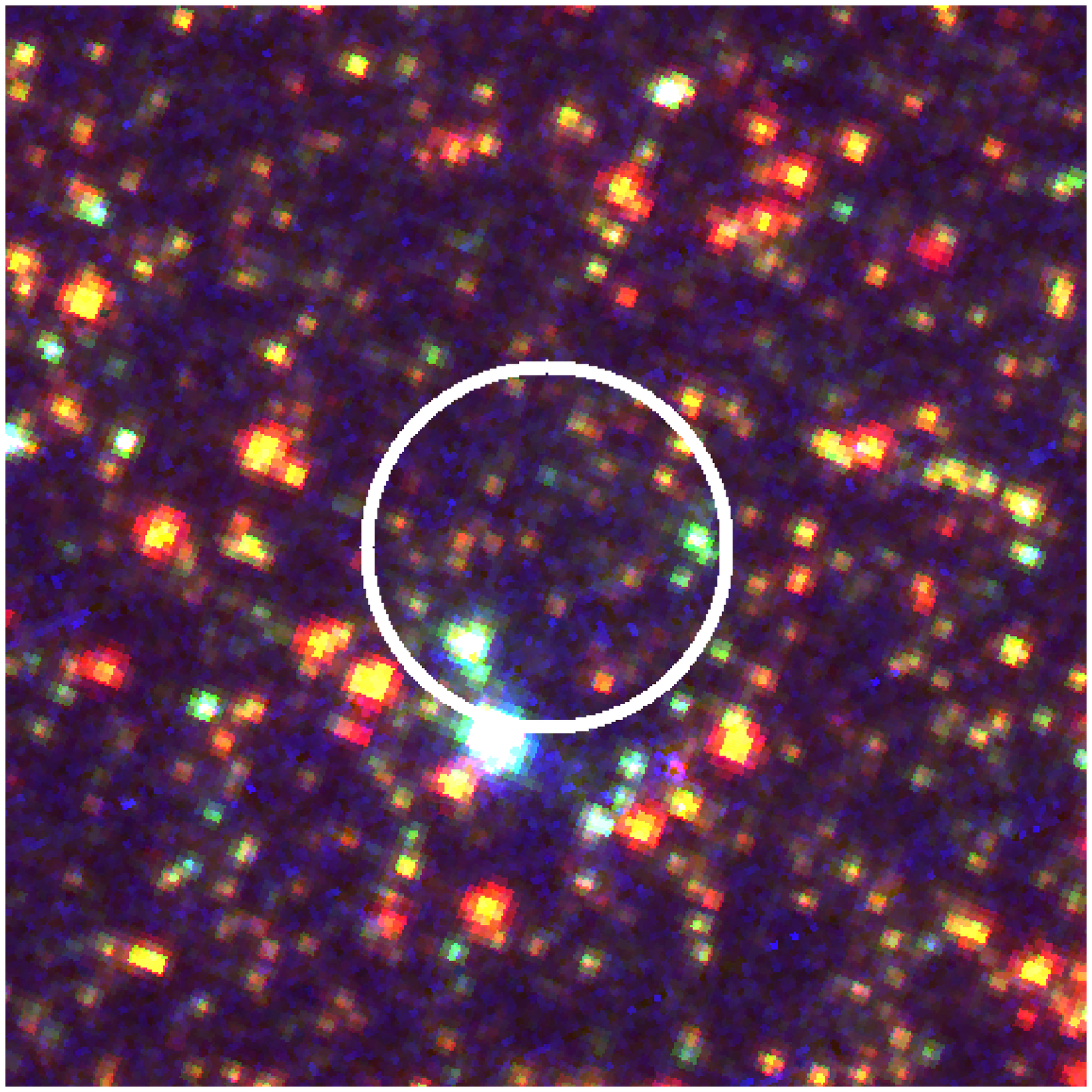}
		\includegraphics[width=\textwidth]{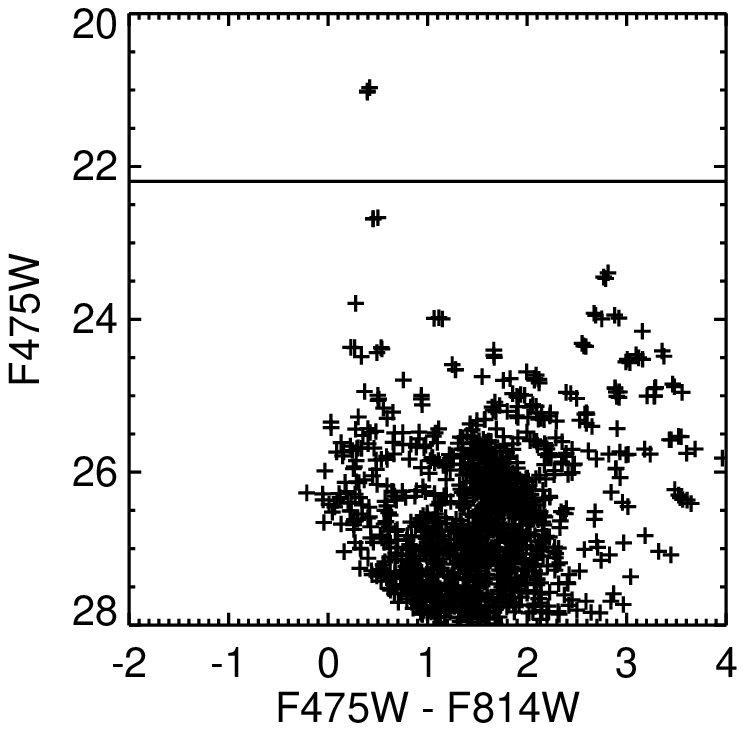}
	\end{minipage}
	\begin{minipage}[t]{.3\textwidth}
		\centering
		\includegraphics[width=\textwidth]{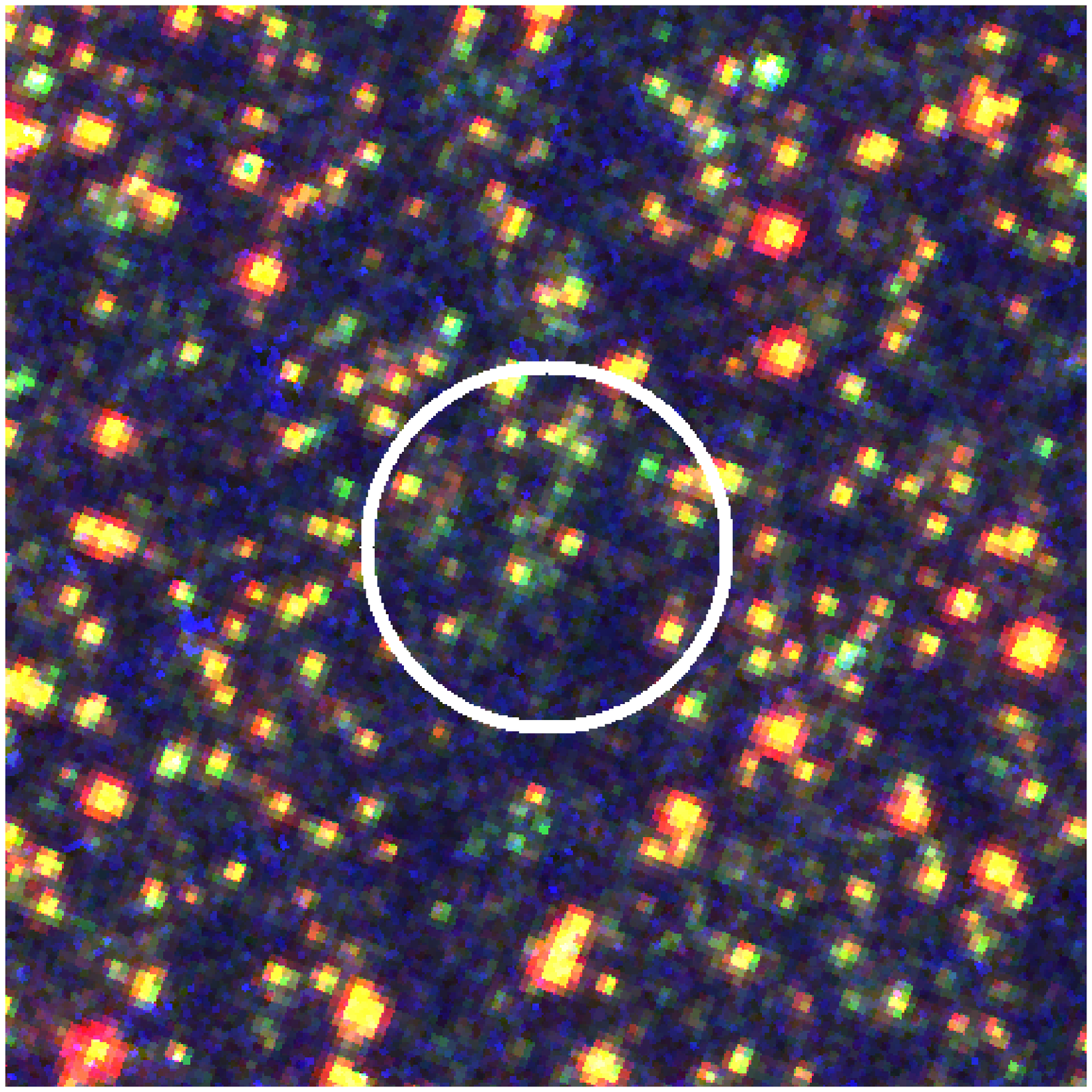}
		\includegraphics[width=\textwidth]{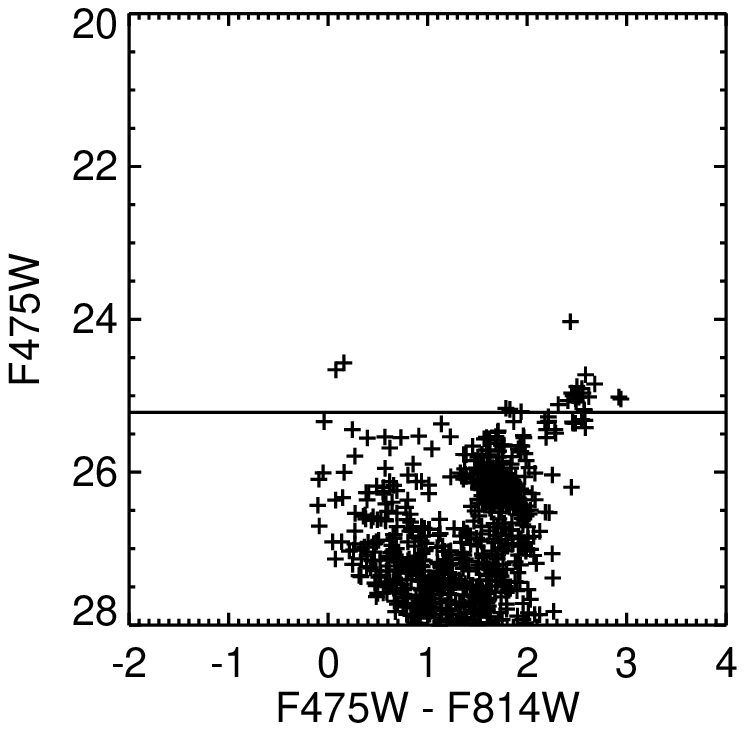}
	\end{minipage}
	\caption{(Above) 3-color images of 3 M06 PN locations where no PN candidate could be
	determined in PHAT. A large white circle with a radius of $1\arcsec$ is centered at the
	M06 RA and Dec for each PN. (Below) The respective optical CMD of stars within $3\arcsec$
	of the M06 PN location. A black line denotes the expected F475W magnitude estimated from
	the M06 $m5007$. From left to right the sources are M06 2721 (a large, bright \ion{H}{2}
	region), M06 1378 (a possible Wolf Rayet star), and M06 731 (no PN candidate in PHAT).}
	\label{nonpnpostage}
\end{figure}

The visual catalog also allowed for the quick identification of possible misidentifications in
the M06 catalog. Locations where there was no probable PN candidate in PHAT and resolved
nebulosity in either the PHAT image or in the LGS [\ion{O}{3}] image were marked as probable
\ion{H}{2} regions. Locations where there was no probable PN candidate in PHAT, no obvious
nebulosity, clear V-band detection, and a single bright main sequence star present in the CMD
with F475W much brighter than expected from $m5007$ were marked as possible stellar emission
line sources, many of which may be Wolf-Rayet stars. Both \ion{H}{2} regions and WR stars can
have strong [\ion{O}{3}] $\lambda$5007 lines, and could have been detected by M06 as [\ion{O}
{3}] emission objects. The left two panels of Figure~\ref{nonpnpostage} show PHAT images of
likely M06 misidentifications - one marked as an \ion{H}{2} region, the other as a stellar
source. A catalog of these possible misidentifications is presented in Section~\ref{results}.
Of the 711 PNe in the M06 catalog that overlap the PHAT images, 152 were likely misidentified.

\begin{figure}
\centering
\includegraphics[width=4in]{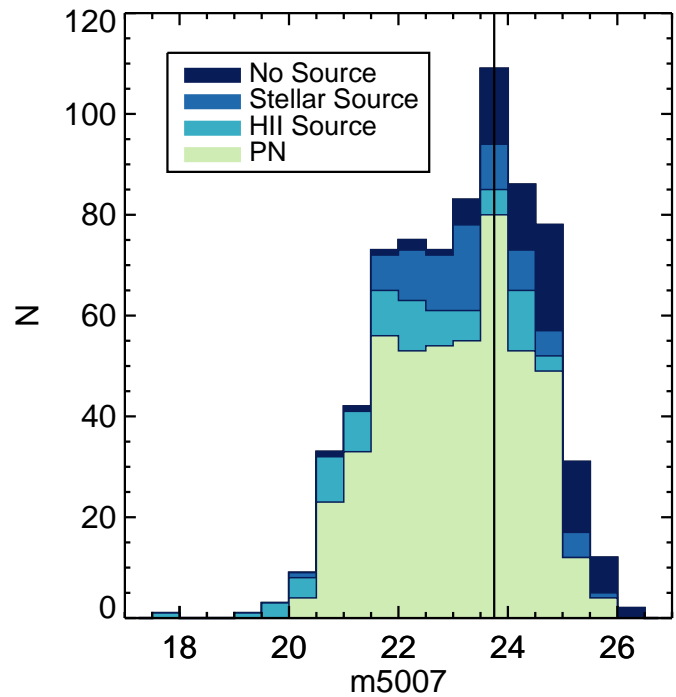}
\caption{Histograms of visually identified sources for all 711 M06 PNe that fall under the
PHAT footprint. The faint end is dominated by fields where no PN could be identified in PHAT
and there were no obvious sources of possible misdetection. A black line is drawn at the
reported M06 completeness limit of $m5007 = 23.75$.}
\label{completeness}
\end{figure}

There were 92 cases where there were no possible PN candidates ($z_{PN}<3$) and no
obvious source of misidentification in the PHAT or LGS images. The frequency of these cases is
much higher for fainter M06 PNe. For M06 PNe with $m5007$ $>$ 25 mag, over half were found to
have no PN candidate in PHAT and no source of misidentification. The right panel of
Figure~\ref{nonpnpostage} shows a PHAT image of an M06 PN location with no PN candidate in
PHAT. These cases are left out of all catalogs and figures. Figure~\ref{completeness} shows
histograms of $m5007$ magnitude for different PHAT classifications.

\section{Results}\label{results}

\begin{figure}
\centering
\includegraphics[width=7in]{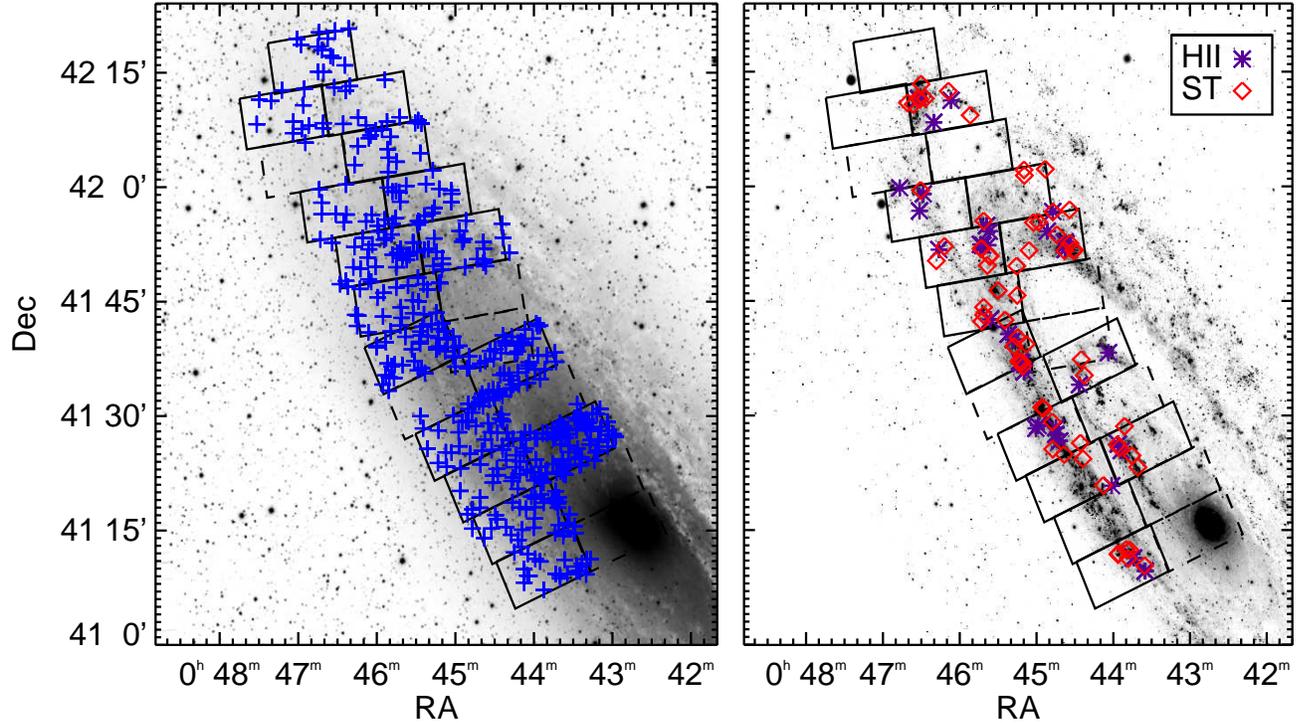}
\caption{(Left) Blue crosses denote the location of matched PNe in the PHAT dataset over an
optical image of M31 (credit and copyright: Martin Pugh
\protect\url{http://www.martinpughastrophotography.id.au/}). The PHAT bricks currently cross
matched with M06 are outlined in solid black lines. The full PHAT footprint is outlined in
dashed black lines. (Right) The location of non-PN sources (extended ``HII'' regions and 
non-PN ``ST''ellar sources) over a GALEX UV image \citep{thilker2005}. PN spatial density
falls off radially while non-PN sources are clustered around UV bright regions. Non-PN objects
were rejected without prior knowledge of their location.}
\label{maps}
\end{figure}

Figure~\ref{maps} shows the spatial distribution of PNe in the PHAT footprint as well as the
locations of possibly misidentified M06 sources. The density of PN sources falls of radially,
tracing the old stellar population. Sources tagged as extended \ion{H}{2} regions or other
non-PN stellar sources are consistently found near regions of recent and on-going star
formation along the disk characterized by high UV flux. Non-PN objects were selected without
prior knowledge of their location.

Table~\ref{pnetable} provides astrometric and photometric measurements in six filters ranging
from near-IR to near-UV for 467 PNe in the footprint of the current PHAT release. Of the 467
PNe in the catalog, 291 have an F275W measurement, 409 have an F336W measurement, 467 have an
F475W measurement, 466 have an F814W measurement, 185 have an F110W measurement, and 156 have
an F160W measurement.

Tables~\ref{hiitable}~\&~\ref{sttable} list M06 PNe for which we found a possible source of
misidentification. These non-PN objects are listed as either extended \ion{H}{2} regions
(Table~\ref{hiitable}) or non-PN stellar sources (Table~\ref{sttable}). Astrometry and 
photometry for stellar sources are included when available. Of these non-PN sources, our 
classifications for 38 sources agree with spectroscopic classification from
\cite{sanders2012}. Most of these are \ion{H}{2} regions.

\subsection{Planetary Nebula Luminosity Function}

The Planetary Nebula Luminosity Function (PNLF) has been proposed as a standard candle for
distance measurements \citep{ciardullo1989}. The PNLF is described by
\begin{equation}\label{pnlfeq}
N(M5007) \propto e^{0.307 \times M5007} \left[ 1 - e^{3 \left( M5007^{*}-M5007 \right)}
\right], 
\end{equation}
where $M5007^{*}$ is the absolute magnitude bright-end cut-off. The same formula can be
applied to apparent magnitude when the distances to the PNe are roughly equivalent, as is the
case with M31. 

\begin{figure}
\centering
\includegraphics[width=4in]{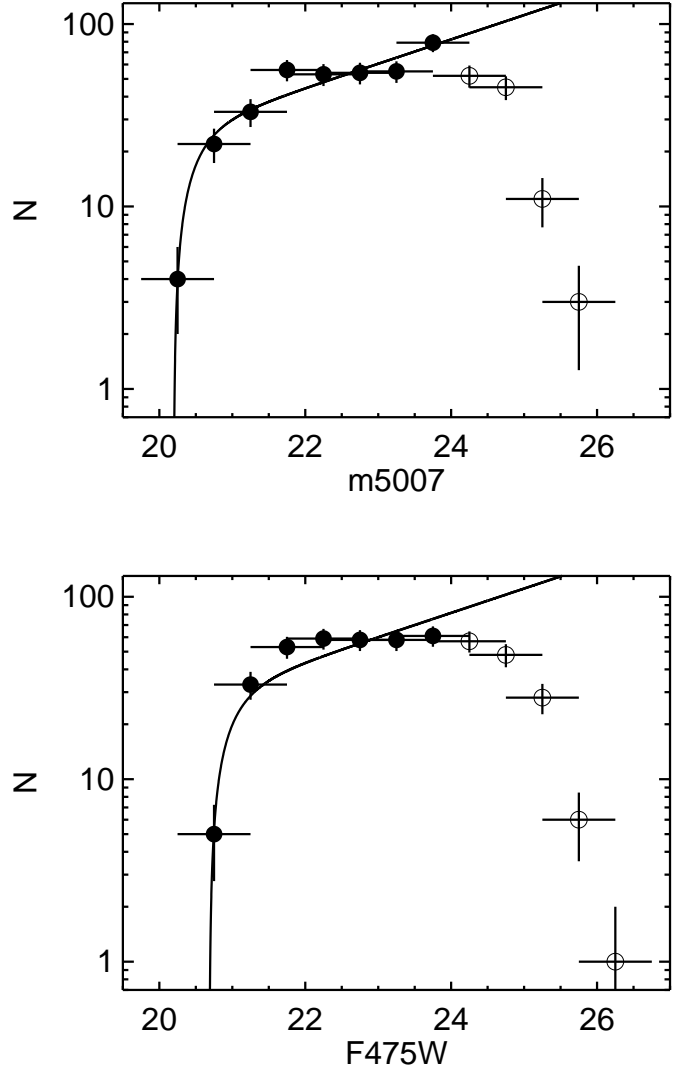}
\caption{The PN luminosity function of the PHAT PNe catalog using $m5007$ (above) and F475W
(below). The solid line shows the best fit to the PNLF as described by Equation~\ref{pnlfeq}.
Filled circles denote values brighter than the completeness limit of 24th magnitude. Open
circles denote values fainter then 24th magnitude which were not used in the fit.}
\label{pnlf}
\end{figure}

The upper half of Figure~\ref{pnlf} shows the PNLF of the PHAT PNe catalog using the M06
$m5007$. The PHAT $m5007$ PNLF matches the expected form up to $\sim$24th magnitude,
consistent with the M06 reported completeness of 23.75. The apparent magnitude bright-end cut
off of the PHAT $m5007$ PNLF is $m5007^{*} = 20.20 \pm 0.03$, which agrees well with the M06
value of $m5007^{*} = 20.2 \pm 0.1$ and the \cite{ciardullo1989} value of $m5007^{*} = 20.17$.

The bottom half of Figure~\ref{pnlf} shows the PHAT PNLF using F475W magnitudes in place of
$m5007$. As one would expect given Figure~\ref{throughput}, the PHAT F475W PNLF is similar in
shape to the $m5007$ PNLF up to $\sim$24th magnitude. There is a slight offset in the apparent
magnitude bright-end cut-off between the $m5007$ and F475W PNLF and the fainter portion of the
F475W PNLF appears flatter. Assuming the F475W PNLF is of the same form as the $m5007$ PNLF,
the apparent magnitude bright-end cut-off of the PHAT F475W PNLF is $F475W^{*} = 20.69 \pm
0.03$. Assuming a distance modulus of $(m-M)_{0} = 24.47 \pm 0.07$ (corresponding to a
distance of $785 \pm 25$ kpc) \citep{mcconnachie2005}, the absolute magnitude bright-end cut
off of the PHAT F475W PNLF is $F475W^{*} = -3.78 \pm 0.08$.

\subsection{CMD Distributions of the PHAT PNe}

We now describe the distribution of PNe in PHAT CMDs; in Section~\ref{evo} we discuss the
evolution of PNe through the CMDs.

\subsubsection{Optical}
\begin{figure}
\centering
\includegraphics[width=5in]{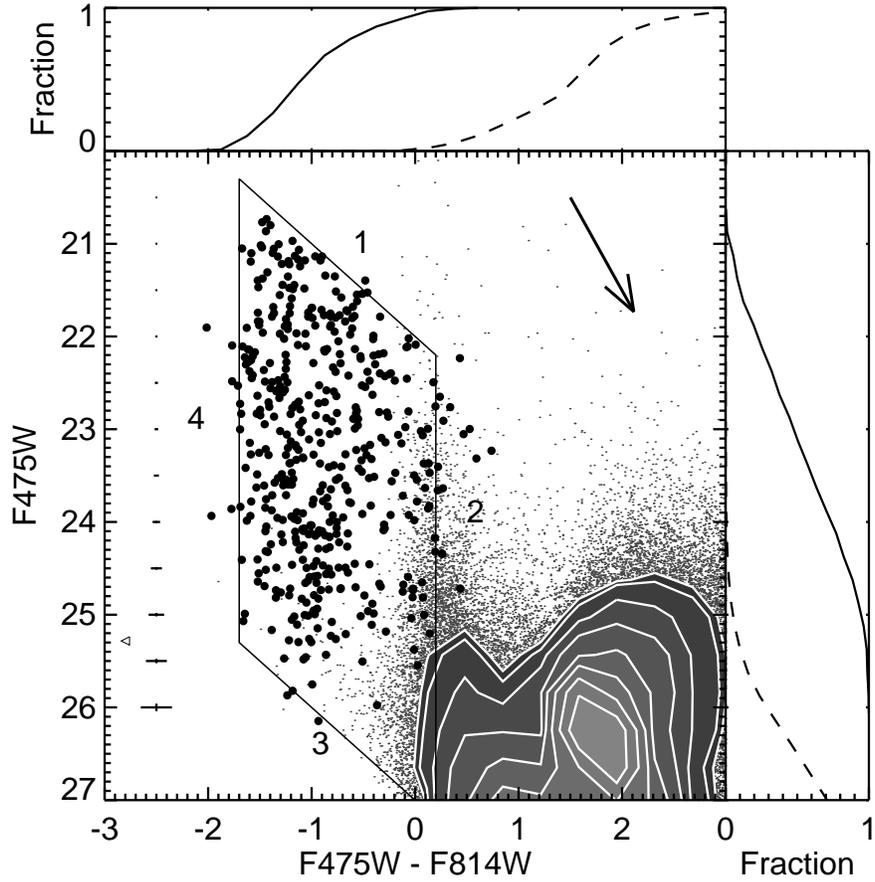}
\caption{Black circles denote PNe. Background grey dots and contours indicate stars within
$3\arcsec$ of a PN. An open triangle at (-2.8, 25.3) represents the F475W magnitude of the one PN with F814W SNR $<$ 4. A rhomboid with sides labeled 1, 2, 3, and 4 outlines the boundaries of
the PNe population. Crosses along the left side of the plot indicate median Poisson errors
from the photometry of PNe binned by F475W. A reddening vector shows the extinction effect of
$A_{V}=1$. (Above) The normalized cumulative distribution of F475W-F814W for PNe (solid line)
and stars within $3\arcsec$ of a PN (dashed line). (Right) The normalized cumulative
distribution of F475W for PNe (solid line) and stars within $3\arcsec$ of a PN (dashed line).
PNe exist in a unique bright-blue region of PHAT optical CMDs.}
\label{opticalcmd}
\end{figure}

An optical CMD of PNe compared with nearby stars (Figure~\ref{opticalcmd}) confirms previous
expectations that most PNe exist in a very blue portion of the CMD where not many stars are
expected to be found. In general, PNe are distinctly brighter and bluer than neighboring stars
($<$3$\arcsec$) as expected from their emission-line dominated spectrum; however, there is a
population of PNe that overlap with main sequence stars in Figure~\ref{opticalcmd}. PNe exist
in a distinct region of the optical CMD as outlined by the parallelogram in
Figure~\ref{opticalcmd}, with hand-drawn boundaries $-1.8 < F475W-F814W < 0.2$ and $22 > F814W
< 27$, containing 95\% of our PNe sample.

\subsubsection{UV}
\begin{figure}
\centering
\includegraphics[width=6in]{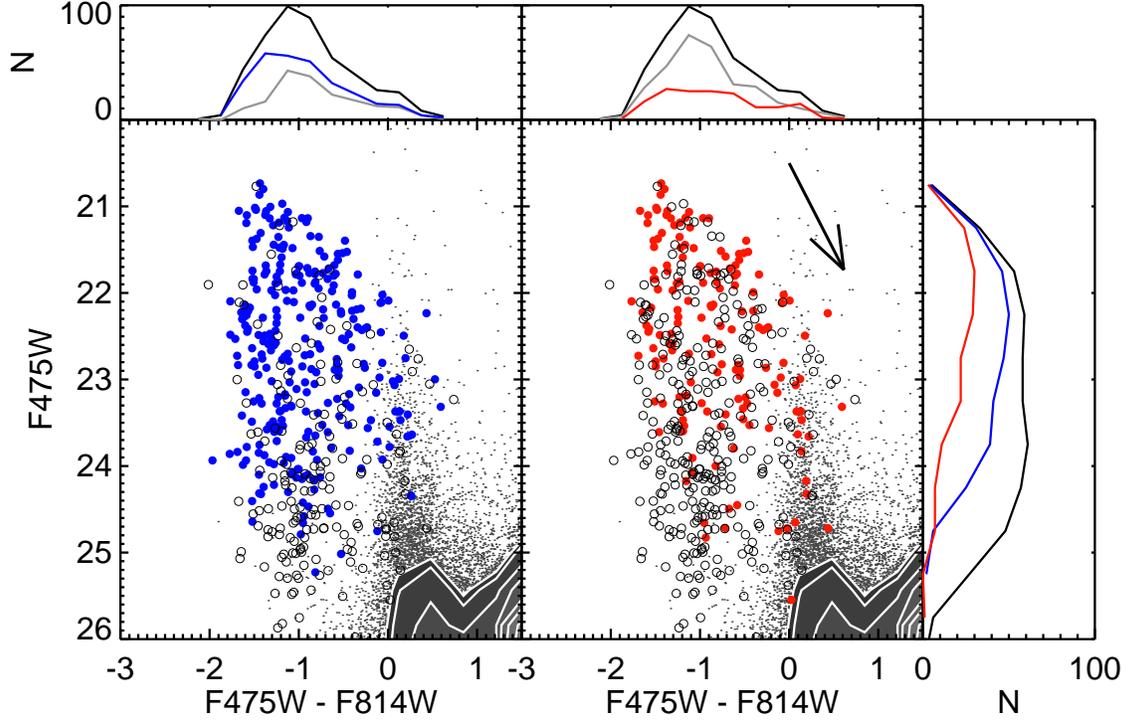}
\caption{(Bottom, Left) Filled, blue circles denote PNe with detections in both UV filters.
Open circles denote PNe without detections in both UV filters. Background grey dots and
contours indicate all stars within $3\arcsec$ of a PN. (Bottom, Center) Filled, red circles
denote PNe with detections in both NIR filters. Open circles denote PNe without detections in
both NIR filters. Background grey dots and contours indicate all stars within $3\arcsec$ of a
PN. A reddening vector shows the extinction effect of $A_{V}=1$. (Top, Left) Distributions of
F475W-F814W for all PNe (black line), PNe with UV detection (blue line), and PNe without UV
detection (grey line). (Top, Right) Distributions of F475W-F814W for all PNe (black line), PNe
with NIR detection (red line), and PNe without NIR detection (grey line). (Bottom, Right)
Distributions of F475W for all PNe (black line), PNe with UV detection (blue line), and PNe
with NIR detection (red line). UV and NIR detection is more common for PNe bright in F475W.}
\label{filterscmd}
\end{figure}

Over $87\%$ of PNe in the PHAT catalog have detections in F336W and $62\%$ have detections in

both UV filters. The left side of Figure~\ref{filterscmd} shows the locations on the optical
CMD of PNe with and without UV detections. UV detection is much more common among PNe that are
bright in F475W, as would be expected for flux due to stellar continuum with some contribution
from emission lines. UV detection is also slightly more common among bluer PNe, as the
brightest PNe also tend to be the bluest.

\begin{figure}
\centering
\includegraphics[width=4in]{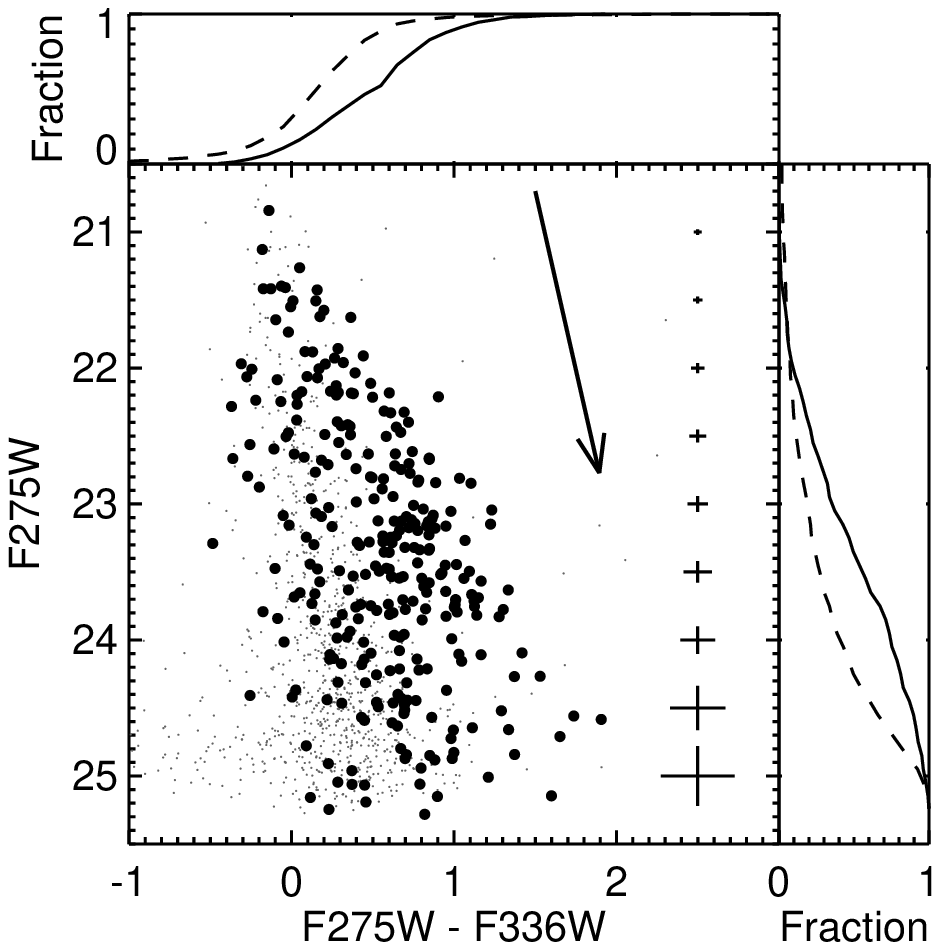}
\caption{Filled, black circles denote PNe. Background grey dots indicate stars within
$3\arcsec$ of a PN. Crosses along the right side of the plot indicate median Poisson errors
from the photometry of PNe binned by F275W. A reddening vector shows the extinction effect of
$A_{V}=1$. (Above) The normalized cumulative distribution of F275W-F336W for PNe (solid line)
and stars within $3\arcsec$ of a PN (dashed line). (Right) The normalized cumulative
distribution of F275W for PNe (solid line) and stars within $3\arcsec$ of a PN (dashed line).
There are only slight differences between PNe and the population of hot MS stars in the PHAT
UV CMD.}
\label{uvcmd}
\end{figure}

Figure~\ref{uvcmd} shows the UV CMD of the PNe with detections in both UV filters. PNe tend to
be brighter and redder in the UV than other stars within $3\arcsec$. However, these
neighboring stars are dominated by MS stars and there is only a slight color difference
between PNe and the population of hot MS stars, as would be expected from the assumption that
UV flux of PNe comes from the very hot central star. The brightest PNe have well constrained
colors, while the spread in color increases for fainter F275W magnitudes which tend to be the
noisiest measurements in the PHAT filter set. 

\subsubsection{NIR}
Over $33\%$ of PNe in our catalog have detections in both NIR filters. Nearly all PNe with

detections in the NIR also have detection in F336W, $84\%$ of which have detections in both UV

filters. Thus, any PN bright enough to be detected in the NIR is almost certainly also
detected in the UV, where proportionally more of the PN's bolometric flux is emitted. The
right side of Figure~\ref{filterscmd} shows the locations on the optical CMD of PNe with and
without NIR detections. Similar to the UV, NIR detection is much more common among PNe that
are bright in F475W. However, the optical color distribution of PNe with detections in NIR is
relatively flat and distinctly different from the optical color distribution of all PNe.

\begin{figure}
\centering
\includegraphics[width=4in]{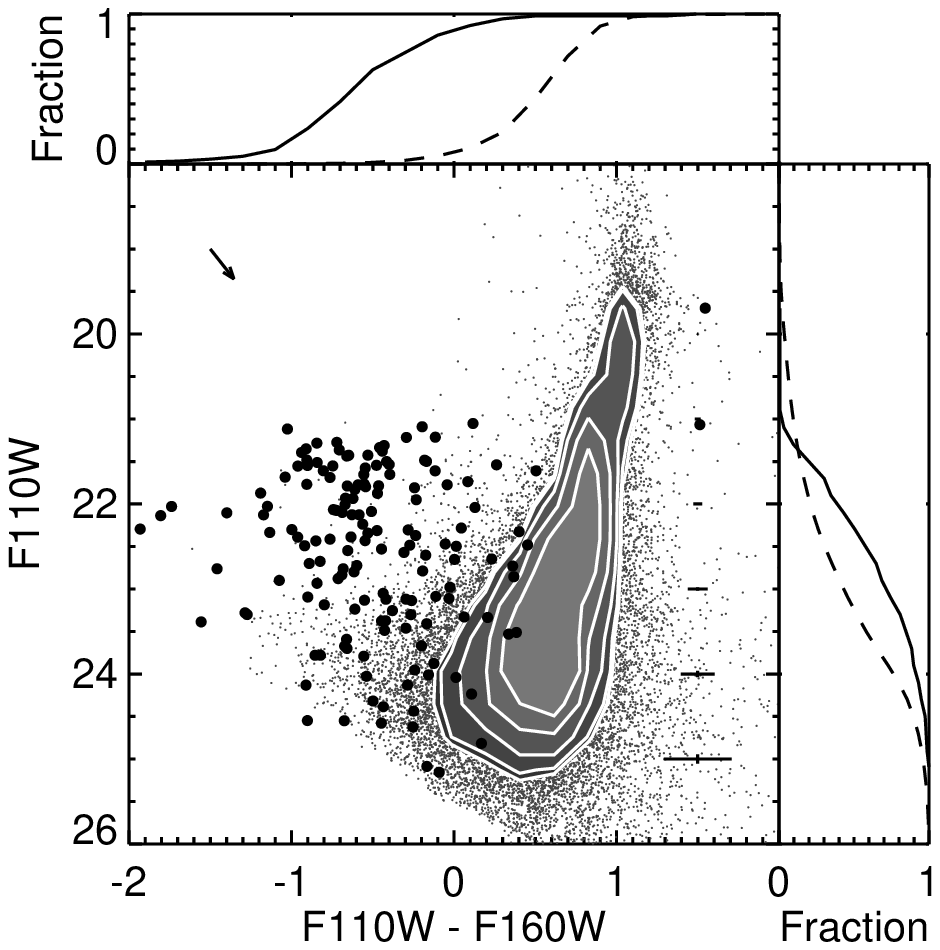}
\caption{Filled, black circles denote PNe. Background grey dots and contours indicate stars
within $3\arcsec$ of a PN. Crosses along the right side of the plot indicate median Poisson
errors from the photometry of PNe binned by F110W. A reddening vector shows the extinction
effect of $A_{V}=1$. (Above) The normalized cumulative distribution of F110W-F160W for PNe
(solid line) and stars within $3\arcsec$ of a PN (dashed line). (Right) The normalized
cumulative distribution of F110W for PNe (solid line) and stars within $3\arcsec$ of a PN
(dashed line). PNe are bluer and brighter in the NIR than neighboring cool RGB and AGB stars.
}
\label{ircmd}
\end{figure}

Figure~\ref{ircmd} shows the NIR CMD of the PNe with detections in both NIR filters. PNe are
distinctly bluer and brighter in the NIR than stars around them, which are primarily cool RGB
and AGB stars. Their distinct color in the NIR is due to the few emission lines in F110W and
the lack of emission lines in F160W. 

There are two NIR-bright PNe in the PHAT catalog with $F110W-F160W > 1$ that are separate from
the rest of the PNe population. It is likely that the PN in each case is coincident with an
NIR-bright source.

\begin{figure}
\centering
\includegraphics[width=4in]{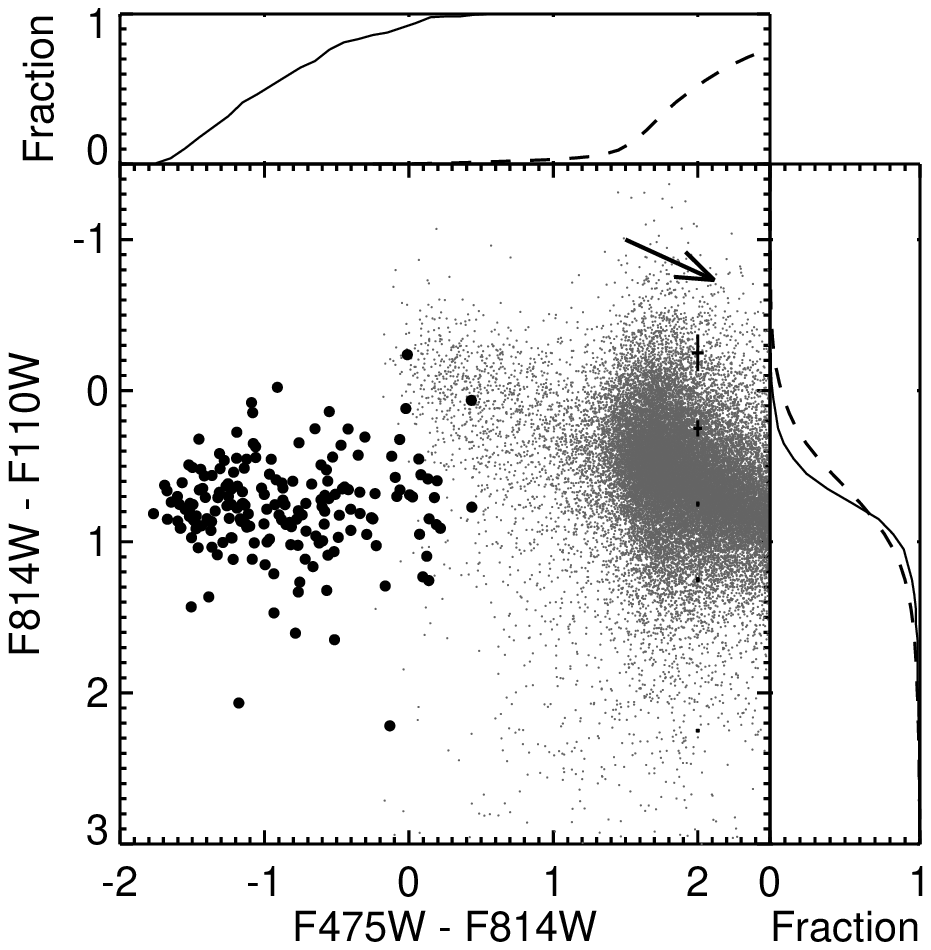}
\caption{Filled, black circles denote PNe. Background grey dots indicate stars
within $3\arcsec$ of a PN. Crosses along the right side of the plot indicate median Poisson
errors from the photometry of PNe binned by F814W-F110W. A reddening vector shows the extinction
effect of $A_{V}=1$. (Above) The normalized cumulative distribution of F475W-F814W for PNe
(solid line) and stars within $3\arcsec$ of a PN (dashed line). (Right) The normalized
cumulative distribution of F814W-F110W for PNe (solid line) and stars within $3\arcsec$ of a PN
(dashed line). PNe are best distinguished from stars by their optical color.
}
\label{optirccd}
\end{figure}

PNe are indeed unique in their photometric signal across all of PHAT's six bands---particularly their optical and NIR color. However, their most distinguishing feature by far is a booming F475W signal best probed by F475W-F814W color. Like in F475W, emission lines in F110W add additional flux that give PNe distinct NIR colors. However, as seen in Figure~\ref{optirccd}, PNe are best secluded by their optical color. Any diagnostic that used multiple bands to identify PNe would necessarily have most of its weight on F475W, as this is where the ratio of line emission to stellar continuum is largest (see Section~\ref{sedsect}).

\subsection{Excitation Classification}
Since PNe are emission line objects, we assume the majority of F475W flux is from emission
lines, allowing us to estimate line flux ratios from the difference between the F475W and
$m5007$ magnitudes. As seen in Figure~\ref{throughput}, the three strongest emission lines
that fall under F475W are [\ion{O}{3}] $\lambda$5007, [\ion{O}{3}] $\lambda$4959, and
$H\beta$. Because the [\ion{O}{3}] $\lambda$5007 to [\ion{O}{3}] $\lambda$4959 ratio is nearly
constant for all PNe, differences in the $m5007-F475W$ color of our PNe are largely due to
differences in the [\ion{O}{3}] to $H\beta$ ratio. This ratio is a measure of ionization in
the nebula and, in general, is expected to be large when the central star is very hot.

\begin{figure}
\centering
\includegraphics[width=4in]{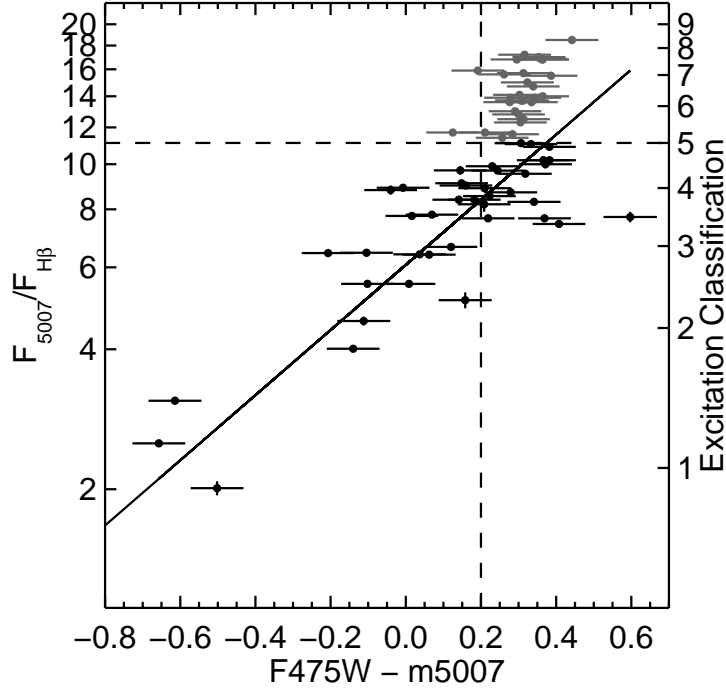}
\caption{The relationship between the difference of F475W and $m5007$ magnitudes and the
$F_{5007}/F_{H\beta}$ line ratio and Excitation classification
($EC\equiv0.45(F_{5007}/F_{H\beta})$, for low EC PNe) using emission line flux ratios from
\cite{sanders2012}. Black points denote low EC PNe ($EC < 5$), for which $H\beta$ makes a
significant contributions to F475W flux. Grey points denote medium to high EC PNe ($EC > 5$),
for which the EC cannot be accurately calculated from $F_{5007}/F_{H\beta}$. A solid line
denotes a linear least squares fit between $F475W - m5007$ and
$log\left(F_{5007}/F_{H\beta}\right)$, ($RMSE = \pm 2.0$). The data has a correlation
coefficient of 0.9. The largest source of error is the uncertainty in $m5007$ of 0.07 mag.
Errors in $F_{5007}/F_{H\beta}$ are often smaller than the plot symbols. Dashed lines indicate
where $F475W-m5007 = 0.2$ and where $EC = 5$. The majority of PNe with $F475W - m5007 < 0.2$
also have $EC < 5$.}
\label{ec}
\end{figure}

To explore the possibility of using $F475W-m5007$ as a probe of the [\ion{O}{3}]$/H\beta$
emission line ratio, we used data from \cite{sanders2012}, who have published spectral line
ratio measurements for several PNe in M31, 69 of which are in the PHAT catalog and have both
$H\beta$ and [\ion{O}{3}] $\lambda$5007 measurements. Figure~\ref{ec} shows the relation
between $F475W-m5007$ and the [\ion{O}{3}] to $H\beta$ ratio, $F_{5007}/F_{H\beta}$. The large
spread of the relation is likely due to differences in the underlying stellar continuum which
can significantly contribute to the total F475W flux when line emission is comparatively weak.
However, a relation between the relative $F475W-m5007$ and $F_{5007}/F_{H\beta}$ exists. For
low values of $F_{5007}/F_{H\beta}$ ($<11$), i.e. comparatively strong $H\beta$ lines, the
relation is quite strong, such that PNe with lower $F_{5007}/F_{H\beta}$ values have brighter
F475W magnitudes in relation to their $m5007$ magnitudes. At higher values of
$F_{5007}/F_{H\beta}$, i.e. comparatively weak $H\beta$ lines, the relation saturates as the
contribution of $H\beta$ to the total flux becomes indistinguishable from the continuum and
weaker emission lines. Therefore, estimates of $F_{5007}/F_{H\beta}$ can only be made for PNe
with $F475W - m5007 < 0.2$. Estimates of the $F_{5007}/F_{H\beta}$ ratio can be made for 170
of the 467 PNe in the PHAT catalog using Equation~\ref{oiiihbeq}.
\begin{equation}\label{oiiihbeq}
F_{5007}/F_{H\beta} = 10^{ 0.7835 + 0.7003 \left( F475W - m5007 \right) }
\end{equation}

The ratio $F_{5007}/F_{H\beta}$ has been used to calculate the Excitation classification (EC)
of low EC PNe ($EC < 5$) \citep{dopita1990,reid2010}. We adopt the $Ex_{\rho}$ method from
\cite{reid2010} for calculating the EC of low EC PNE as defined in Equation~\ref{eceq}.
\begin{equation}\label{eceq}
EC \equiv 0.45 \times (F_{5007}/F_{H\beta})
\end{equation}
EC estimates for the 170 PNe with $F475W - m5007 < 0.2$ are included in Table~\ref{pnetable}
with an uncertainty of $\sigma_{EC} = \pm1$. All PNe with $F475W - m5007 \geq 0.2$ are assumed
to have EC $> 3.8$. As excitation increases, the contributions of other emission lines 
such as \ion{He}{2} become comparable with $H\beta$ which would cause an underestimate in the 
PN's EC.

\section{Discussion}\label{discussion}

\subsection{Extinction}
\begin{figure}
\centering
\includegraphics[width=5in]{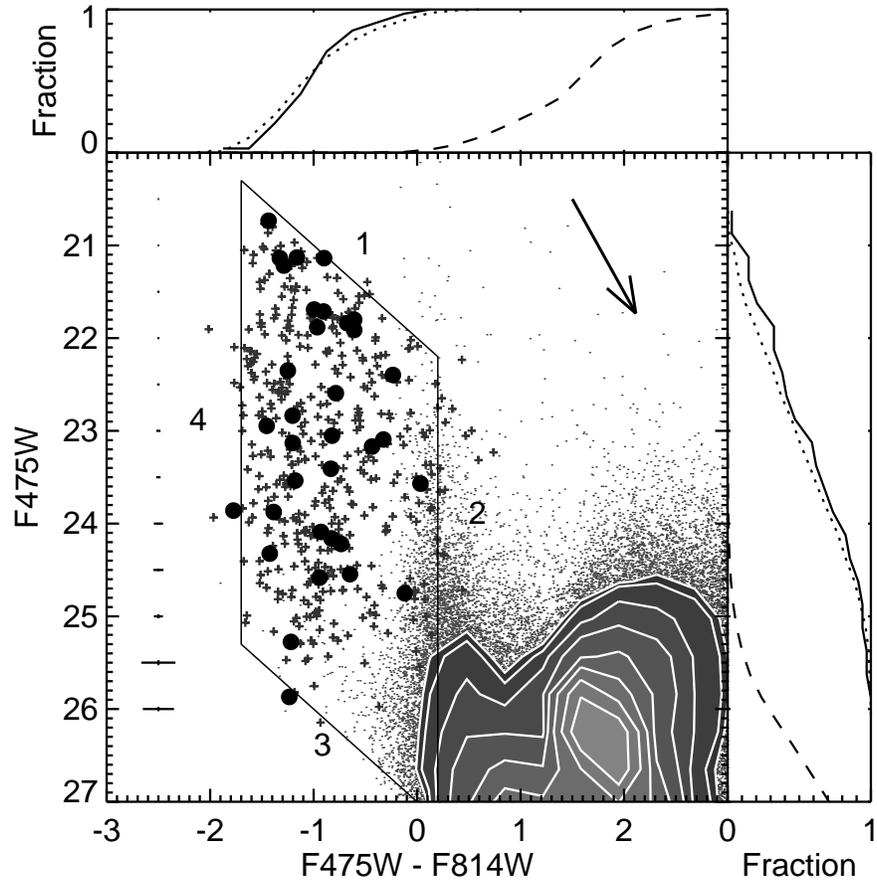}
\caption{Same as Figure~\ref{opticalcmd}, but now large black circles and solid lines 
denote only PNe from ``dust-free'' regions. Grey crosses and dotted lines denote the full 
sample. The distribution is qualitatively very similar to that of the full sample suggesting 
the PHAT PNe catalog does not suffer large differential extinction.}
\label{lowav}
\end{figure}

M31 is known to be a dusty galaxy \citep[e.g.][Dalcanton et al. 2014, submitted]{draine2014}.
Therefore, we need to understand how dust
may be affecting our PNe photometry. To determine the degree to which
the PNe sample is attenuated by extinction, we selected a subsample of PNe from
``dust-free'' regions as determined from the extinction maps of
Dalcanton et al. 2014 (submitted). These PNe were selected to be in
regions with A$_V{<}$0.5 and a low fraction of reddened stars. Their optical CMD is shown in
Figure~\ref{lowav}, and the distribution is qualitatively
equivalent to that of the full sample. Perhaps because PNe tend not
to be associated with ongoing star formation or because PNe are
associated with old stars ($>$ 1 Gyr) that have migrated out of the galactic
plane, our PNe photometry appears to be relatively unaffected by dust
extinction.

\subsection{PNLF}
The consistency of the PHAT $m5007$ PNLF with the M06 PNLF suggests that the PHAT PNe catalog
is a representative and unbiased subset of the M06 catalog. Despite the fact that the F475W
bandpass covers emission lines other than [\ion{O}{3}] $\lambda$5007 and continuum from the
central star, the PHAT F475W PNLF is similar in form to the $m5007$ PNLF, particularly on the
bright-end -- aside from a slight offset. This is because the contribution by stellar
continuum and weaker emission lines to the total F475W flux is less significant in brighter
PNe, for which the F475W magnitude is a good indicator of $m5007$, making the F475W PNLF a
possible proxy for the traditional $m5007$ PNLF for future extragalactic surveys. The flatness
of the faint-end of the PHAT F475W PNLF could be due to the larger relative contribution by
stellar continuum for PNe with weaker [\ion{O}{3}] emission.

\subsection{Excitation Classification and PN Evolution}\label{evo}
Excitation classification has been found to depend on the evolutionary state of a PN, which is
reflected in the central star's effective temperature, the PN's radius, and the PN's expansion
velocity \citep{dopita1987, dopita1988, dopita1990, dopita1991i, dopita1991ii, gurzadyan1991,
reid2010}. Because many parameters affect the EC, the correlations between EC and physical
properties of PNe have large scatter which, coupled with the large uncertainties of our own EC
estimates, means we cannot reliably estimate individual parameters. However, our ability to
isolate a subsample of low EC PNe allows us to place some PNe in the evolutionarily young
phase with small sizes and low central star temperatures. For the remainder of this paper, we
referred to PNe with EC below the upper limit of our ability to determine EC ($EC < 3.8$) as
low EC PNe and all other PNe as medium to high EC PNe; note that the canonical definition of
``low EC'' includes up to an EC of 5. We do expect some cross contamination due to the large
uncertainties in our $F475W - m5007$ measurements such that some PNe in our sample of low EC
PNe are in fact medium to high EC PNe and vice versa. In addition, some high EC PNe may 
be misclassified as low EC PNe due to increased contribution of \ion{He}{2} and other weak 
lines in the F475W bandpass.

According to PN evolutionary tracks based on photoionization models described in
\cite{dopita1990} and \cite{reid2010}, all PNe, regardless of their initial mass, begin as low
EC PNe. They then increase in EC as they evolve to a mass-dependent maximum EC before
returning to a slightly lower EC. Only lower mass PNe (central star stellar mass
$\lesssim$0.6$M_{\odot}$) are expected to return to an EC lower than 5. The majority of low EC
PNe in our catalog are therefore likely to be young, but with a wide range of initial masses.
Additionally, high EC PNe in our catalog are expected to cover the entire range of PNe ages
except for the youngest.

Another set of PN evolutionary models useful for understanding how PNe evolve through PHAT
color-magnitude and color-color diagrams are 1D radiation-hydrodynamic models that describe
the evolution of PNe by modeling a circumstellar envelope along with the central star as it
evolves from the AGB toward the white-dwarf cooling path \citep{perinotto2004}. Studies of the
emission properties of these models \citep[e.g.][]{schonberner2007,mendez2008} explain that in
early stages of evolution, PNe increase in $m5007$ brightness as they increase in EC until
they reach a maximum $m5007$ brightness. The intensity and duration of maximum m5007 brightness is mass-dependent. In general, higher mass progenitors result in brighter maximum m5007 but for shorter duration than fainter low mass progenitors. \citep{kwok2000}. Maximum $m5007$ brightness is reached at ECs higher
than the upper limit of our ability to determine EC of PNe in PHAT. Therefore, we expect the
PNe in our catalog to evolve from low EC to high EC simultaneously increasing in $m5007$
brightness to a maximum brightness. Indeed, this is the qualitative behavior observed in
Figure~\ref{opticalcmdec}, as described below.

\subsubsection{PN Evolution Through CMDs}
\begin{figure}
\centering
\includegraphics[width=5in]{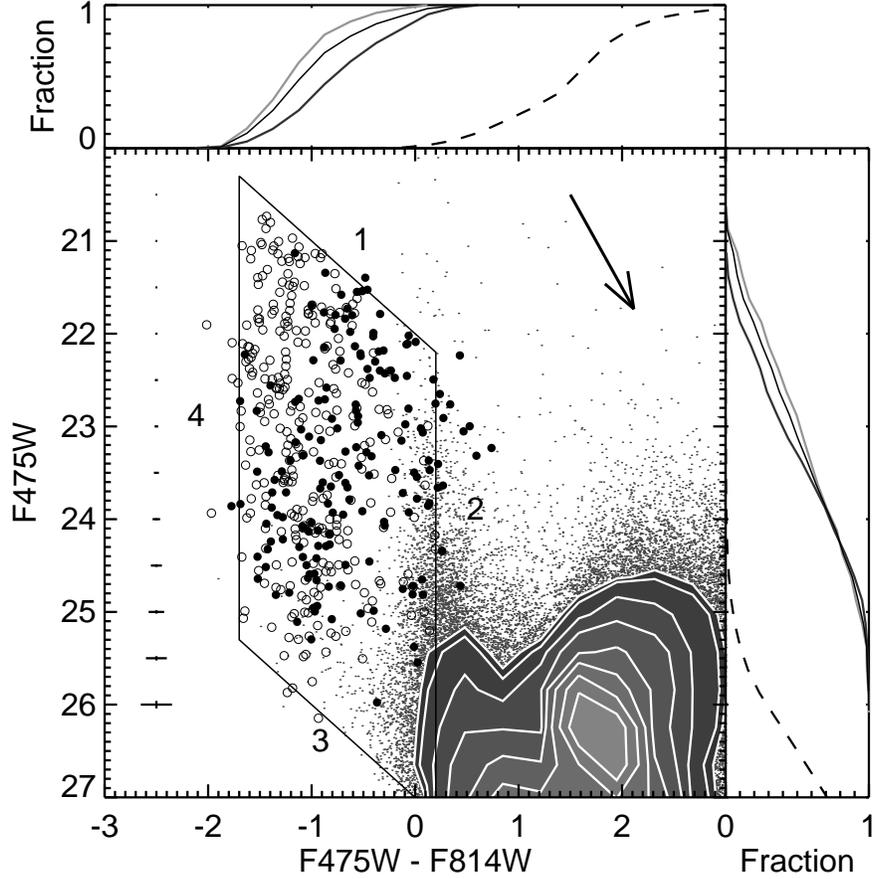}
\caption{Filled, black circles denote low EC PNe. Open circles denote medium to high EC PNe.
Background grey dots and contours indicate stars within $3\arcsec$ of a PN. A rhomboid with
sides labeled 1, 2, 3, and 4 outlines the boundaries of the PNe population. Crosses along the
left side of the plot indicate median Poisson errors from the photometry of PNe binned by
F475W. A reddening vector shows the extinction effect of $A_{V}=1$. (Above) The normalized
cumulative distribution of F475W-F814W for PNe (thin black line), low EC PNe (dark grey line),
medium to high EC PNe (light grey line), and stars within $3\arcsec$ of a PN (dashed line).
(Right) The normalized cumulative distribution of F475W for PNe (thin black line), low EC PNe
(dark grey line), medium to high EC PNe (light grey line), and stars within $3\arcsec$ of a PN
(dashed line). There is a distinct bright, blue population of medium to high EC PNe suggesting
an evolutionary transition to higher F475W flux as increasing central star temperature
increases excitation in the nebula.}
\label{opticalcmdec}
\end{figure}

Figure~\ref{opticalcmdec} shows the locations of low EC and medium to high EC PNe on an
optical CMD. We see a bifurcation of low EC and medium to high EC PNe among PNe with F475W $<
23.5$. This behavior would be consistent with the evolutionary transition of a high mass progenitor along, or parallel
to, Side 1 (the upper boundary in the figure) as the PN's [\ion{O}{3}] $\lambda$5007 flux
reaches a maximum. The proposed transition aligns well with the PN evolutionary models
described above, namely a transition from low EC to higher EC and from fainter F475W
magnitudes to the brightest. Brightening in F475W along Side 1 toward the bluer portion of the
CMD corresponds to evolution with a constant F814W magnitude.

The less distinct red boundary indicated by Side 2 in Figure~\ref{opticalcmdec} is made up of
mostly low EC PNe and overlaps main sequence stars. The overlap with the main sequence is most
likely due to interstellar reddening of young PNe but could also indicate residual
contamination from WR stars. The faint boundary (Side 3) and the similar boundary for field
stars are products of the cuts in signal-to-noise.

\begin{figure}
\centering
\includegraphics[width=4in]{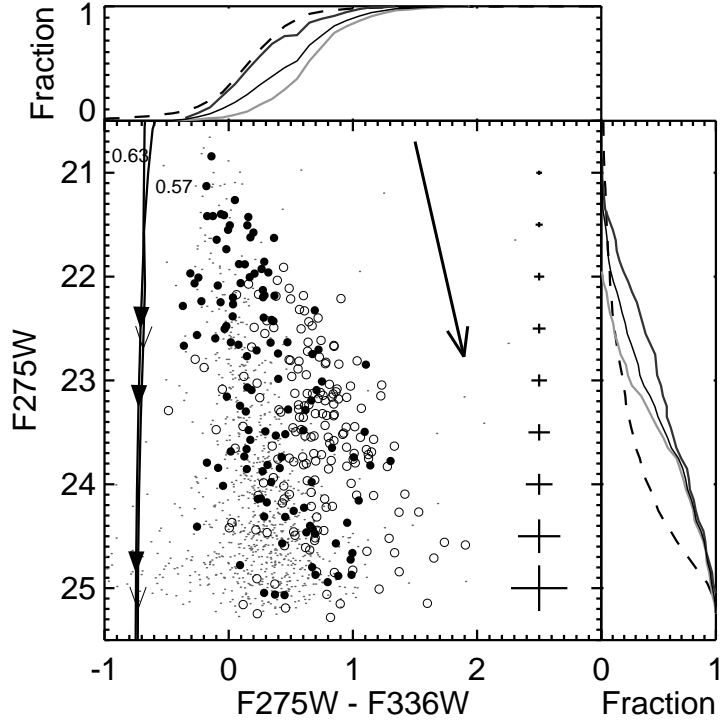}
\caption{Filled, black circles denote low EC PNe. Open circles denote medium to high EC PNe.
Background grey dots indicate stars within $3\arcsec$ of a PN. Crosses along the left side of
the plot indicate median Poisson errors from the photometry of PNe binned by F275W. A
reddening vector shows the extinction effect of $A_{V}=1$. Also shown are two unreddened P-AGB
tracks for initial masses of $0.57M_{\odot}$ and $0.63M_{\odot}$. The tracks are nearly identical in this region of color-magnitude space. Open arrows indicate 
$0.57M_{\odot}$ 20,000 and 30,000 year ages. Filled arrows indicate $0.63M_{\odot}$ 2,000, 3,000, and 
4,000 ages. (Above) The normalized cumulative distribution of F275W-F336W for PNe (thin black line), 
low EC PNe (dark grey line), medium to high EC PNe (light grey line), and stars within $3\arcsec$ of 
a PN (dashed line). (Right) The normalized cumulative distribution of F275W for PNe (thin black 
line), low EC PNe (dark grey line), medium to high EC PNe (light grey line), and stars within 
$3\arcsec$ of a PN (dashed line). Low EC PNe tend to be brighter and bluer than medium to high EC PNe 
in the PHAT UV CMD.}
\label{uvcmdec}
\end{figure}

Figure~\ref{uvcmdec} shows the locations of low EC and medium to high
EC PNe on a UV CMD. The low EC PNe are systematically brighter and bluer, suggesting some
transition toward redder and fainter UV magnitudes
as they evolve into higher ECs. The slightly redder extant of the PN distribution compared to
bright MS stars is likely due to increased contribution from nebular continuum in the F336W
bandpass. Also shown in Figure~\ref{uvcmdec} are two Post-AGB (P-AGB) evolutionary tracks from 
\cite{vassiliadis1993} converted to WFC3/UVIS photometry as described in \cite{girardi2008},
shown without extinction correction. We see an increased spread of UV color at fainter F275W
magnitudes, but it is hard to say for certain if the spread represents the true spread in PNe
UV color, or is due to photometric uncertainty and/or interstellar reddening. Still, it is
clear that reduced UV flux is a prominent difference between low EC and medium to high EC PNe
on the UV CMD. Such a transition is expected from P-AGB evolutionary tracks.

\begin{figure}
\centering
\includegraphics[width=4in]{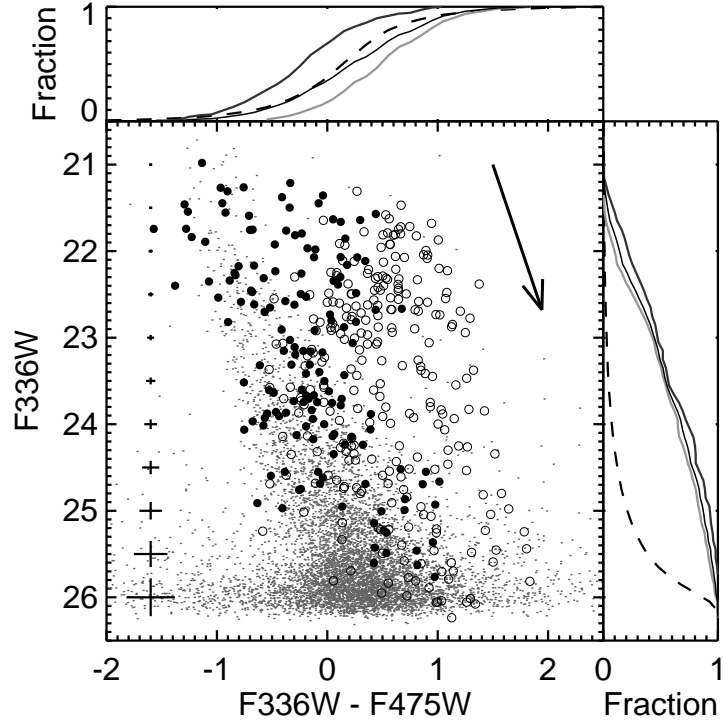}
\caption{Filled, black circles denote low EC PNe. Open circles denote medium to high EC PNe.
Background grey dots indicate stars within $3\arcsec$ of a PN. Crosses along the left side of
the plot indicate median Poisson errors from the photometry of PNe binned by F336W. A
reddening vector shows the extinction effect of $A_{V}=1$. (Above) The normalized cumulative
distribution of F336W-F475W for PNe (thin black line), low EC PNe (dark grey line), medium to
high EC PNe (light grey line), and stars within $3\arcsec$ of a PN (dashed line). (Right) The
normalized cumulative distribution of F336W for PNe (thin black line), low EC PNe (dark grey
line), medium to high EC PNe (light grey line), and stars within $3\arcsec$ of a PN (dashed
line). A separation in excitation becomes apparent when comparing emission flux to continuum
flux.}
\label{uvoptcmdec}
\end{figure}

Figure~\ref{uvoptcmdec} shows the CMD for F336W-F475W vs F336W. Because F336W is dominated by
stellar and nebular continuum and F475W by line emission, this diagram compares emission line
flux to underlying continuum. Low EC PNe appear much like hot stars, but medium to high EC PNe
show up where few stars are expected to be found. The increased excitation increases flux in
the emission line-dominated F475W bandpass, pulling PNe redward to a relatively unpopulated
portion of the CMD. This diagnostic may be useful for future studies of PNe as it is possible
to replicate through ground-based observations. Of the six PHAT filters, PNe are brightest in
F336W and F475W.

In the NIR there is no separation in either color or magnitude between low EC and medium to
high EC PNe. MS stars in PHAT are expected to have NIR colors redder than $F110W - F160W = 0$,
with only younger MS stars ($<$ 100 Myr) expected to have bluer colors at $F110W <\:\sim$23.
The lack of young MS stars neighboring PNe (Figure~\ref{ircmd}) emphasizes that PNe are not
sampling regions with very recent star formation.

\subsection{Spectral Energy Distributions}\label{sedsect}
\begin{figure}
\centering
\includegraphics[width=0.9\textwidth]{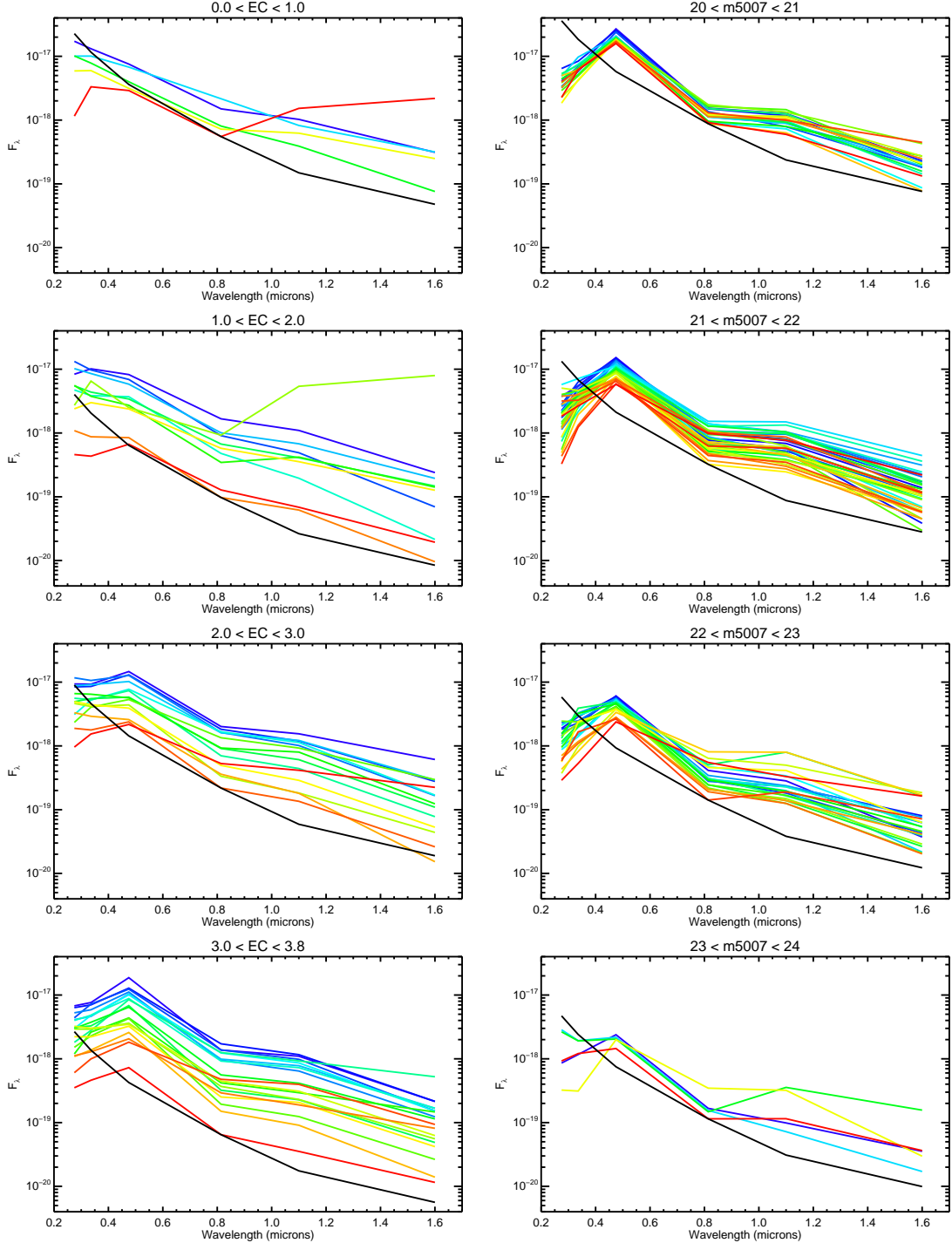}
\caption{Spectral energy distributions of PHAT PNe with detections in all six bandpasses. The
left column of plots shows SEDs of low EC PNe sorted by EC. The right column of plots show
SEDs of medium to high EC PNe sorted by $m5007$. All SED plots are color-ranked by relative
F475W flux. Black lines denote a 50,000 K black body curve accounting for full system
throughput and extinction-corrected for $A_{V} = 0.17$ ($R_{V} = 3.1$), the average foreground
extinction for the PHAT region \citep{schlafly2011}. Increasing excitation correlates to
increasing emission line flux over the underlying continuum in the F475W and F110W
bandpasses.}
\label{seds}
\end{figure}

For PNe with detections in all six PHAT filters, converting from
magnitude to total flux produces Spectral Energy Distributions (SEDs)
covering 0.3 to 1.6 microns. Such SEDs will provide constraints for
future models of PNe evolution. Such models will be necessary to
reliably constrain extinction for these PNe. Figure~\ref{seds} shows
SEDs for 130 PNe with detections in all six filters.

The right column of Figure~\ref{seds} shows SEDs grouped by $m5007$. SEDs of the brightest PNe
are largely uniform while SEDs of fainter PNe show more variations. The SEDs look much like
those of hot stars with the addition of a prominent peaks at 0.475 and 1.1 microns due to 
emission lines in the bandpasses. However, unlike hot stars, flux at 0.336 microns exceeds 
flux at 0.275 microns for many PNe. This could be due to the fact that 0.275 micron flux is
more attenuated by extinction, combined with the larger uncertainties in 0.275 micron flux. 

The left column of Figure~\ref{seds} shows SEDs grouped by EC. At the lowest EC, the SEDs show
little evidence of emission line flux above the blue stellar continuum. Moving through higher
EC, the emission line features become much more prominent. Again, this suggests low EC PNe are
largely young PNe with the lowest EC PNe being just beyond the emergence of [\ion{O}{3}]
lines, which grow stronger as PNe evolve to higher EC.

The two anomalously NIR-bright PNe also show up in the SEDs. Again, it is likely they are
crowded by a NIR-bright source.

\subsection{Resolved PNe and Size Estimates}
\begin{figure}
\centering
\includegraphics[width=4in]{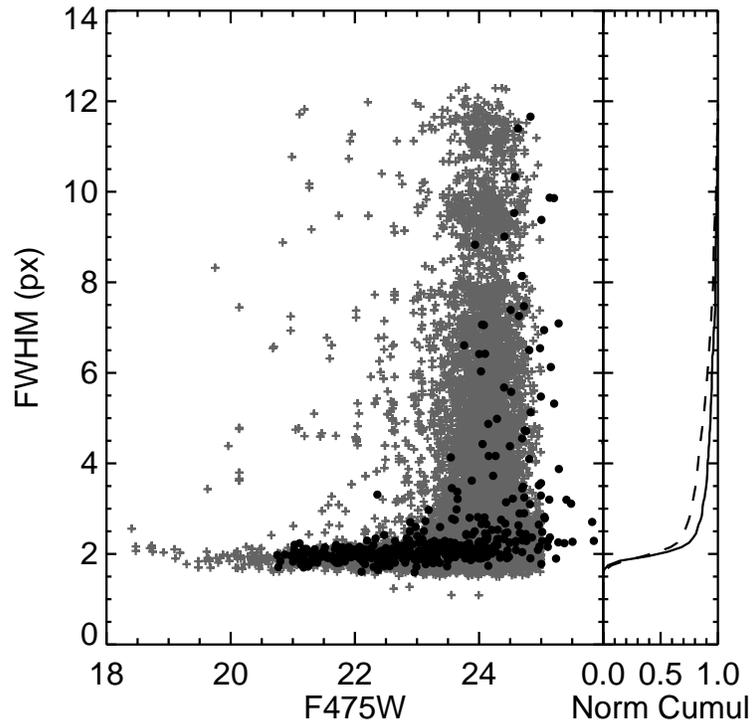}
\caption{(Left) Direct FWHM measurements of PNe (black circles) and the 50 brightest stars
with F475W$<$25 around each PN (grey pluses). FWHM values greater than $\sim$3 are likely due
to poor direct FWHM measurement of faint, crowded sources. (Right) Normalized cumulative
histograms of FWHM for PNe (solid line) and stars (dashed line). The PHAT PNe catalog does not
appear extended compared  to neighboring stars.}
\label{fwhm}
\end{figure}

PNe are commonly found to have radii on the order of 0.1 pc but have been known to have radii
as large as 0.6 pc \citep{phillips2003}. It might be expected that PNe with radii $>$0.5 pc
could be resolved in PHAT considering HST's ACS camera has a plate scale of $0\farcs05$/pixel
(0.2 pc/pixel at the distance of M31) and the average full width at half max (FWHM) of PHAT's
PSF is $\sim$2 pixels. However, as PNe evolve they both expand radially and drop in
luminosity, resulting in a steep decline of surface brightness. PNe large enough to be
resolved in PHAT are likely too faint to be detected in PHAT. At intermediate stages of
evolution, however, some PNe may be bright enough to be detected in PHAT and large enough to
be slightly resolved. 
Figure~\ref{fwhm} shows the dependence of FWHM on magnitude in PHAT data. Faint sources in
crowded regions suffer heavy blending that can cause the direct FWHM measurement to
overestimate the true FWHM. However, the FWHM distribution of PNe is indistinguishable from
that of nearby stars, suggesting that PNe are not resolved in PHAT. 

\begin{figure}
\centering
\includegraphics[width=4in]{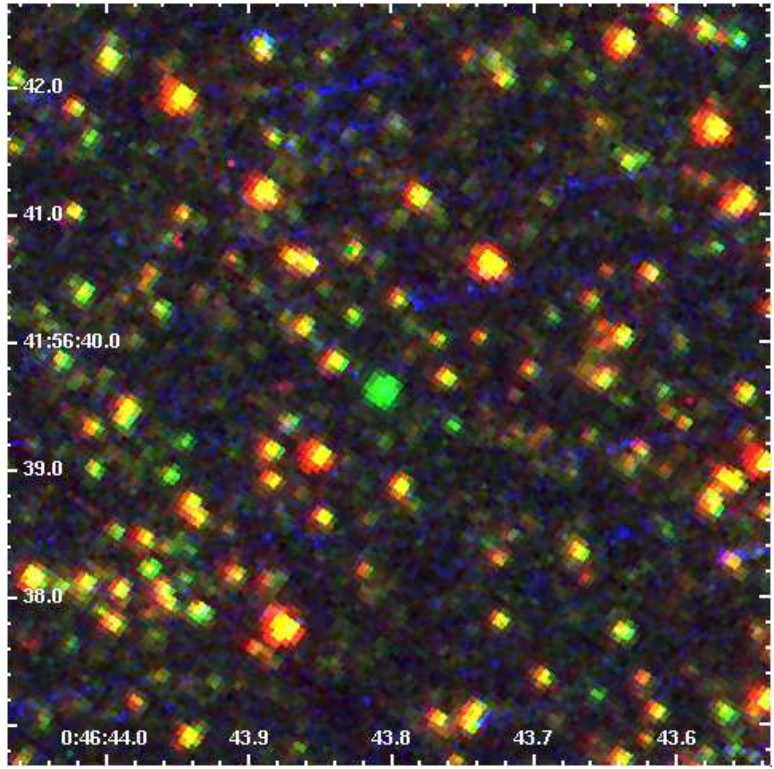}
\caption{The bright, green object in the center of the frame is M06 143 - a slightly resolved
PN candidate.}
\label{143}
\end{figure}

In general, the PNe sample does not appear any more extended than nearby stars, however, we
found one PN (M06 143) that appeared
significantly larger than the surrounding stars. We show an image of
this PN in Figure~\ref{143}. This PN has a FWHM of $\sim$4 pixels,
compared to stars in the area, which have FWHMs of $\sim$2 pixels. After
accounting for the effects of a $0\farcs1$ FWHM point spread function,
the intrinsic radius of the PN would be up to $0\farcs08$, or 0.3$-$0.4 pc. The PHAT
magnitudes of M06 143 are likely underestimated due to the extended nature of the PSF. This is
supported by the fact that M06 143 is the faintest PN in our sample, with $F475W=26.15$, and
has $F475W-m5007=1.4$ --- the largest difference of the PNe sample and more than three times
the $F475W-m5007$ saturation limit of the PNe sample. Though the optical color of M06 143 is
consistent with PNe in PHAT, we cannot rule out the possibility it is a blended source or a
non-PN source.

Finally, any M06 source that was visually resolved in PHAT was rejected as being a PN (M06 143
is the only marginally extended PN) and classified as a possible \ion{H}{2} region. Therefore,
inclusion in Table~\ref{hiitable} serves as an updated classification of extended objects in
M06, which will be more reliable than such classifications from the ground-based data.

\section{Summary}\label{summary}

We have identified 467 PNe in the PHAT survey photometry catalogs.
The F475W broadband magnitudes show a tight correlation with the $m5007$
magnitudes clearly indicating the purity of our sample. These PNe are
unresolved and dominated by [\ion{O}{3}] line emission. Since
the PHAT astrometric solution is precise to $0\farcs01$, we have provided improved astrometry
for all of the PNe in our sample, along with 6-band photometry from the PHAT survey.

Comparing our broadband magnitudes and M06 narrowband magnitudes to excitation classifications
from optical spectroscopy in the literature shows that low excitation PNe
tend to have bluer colors in F475W-$m5007$, suggesting that our
broadband magnitudes in these cases probe the emission from the
central star. These results indicate that our SEDs will be useful for
constraining models of PNe and central star evolution. 

The size distribution of PNe in our HST images are consistent with
that of point sources. However, there is some evidence that at least
one PN is marginally resolved, corresponding to a physical radius of up to
0.3$-$0.4 pc.

\acknowledgments
Support for this work was provided by NASA through grant GO-12055 from
the Space Telescope Science Institute, which is operated by the
Association of Universities for Research in Astronomy, Incorporated,
under NASA contract NAS5-26555.

\clearpage

\clearpage

\clearpage

\clearpage

\footnotesize
\begin{deluxetable}{ll}\tablewidth{0pt}
\rotate
\footnotesize
\tablecaption{Flux Contribution by Filter}
\tablehead{
\colhead{{\footnotesize Filter}} &
\colhead{{\footnotesize Likely Main Flux Contributors}}\\
}
\startdata
F275W & Stellar Continuum, Nebular Continuum, [\ion{Fe}{4}] $\lambda$2836, [\ion{Fe}{4}]
        $\lambda$2829, [\ion{Fe}{4}] $\lambda$2567, [\ion{C}{4}] $\lambda$2529,\\
      & \ion{He}{2} $\lambda$2733, \ion{He}{2} $\lambda$2511, \ion{He}{2} $\lambda$2386,
        [\ion{Ar}{4}] $\lambda$2854\\
F336W & Stellar Continuum, Nebular Continuum, Balmer Continuum, \ion{He}{2} $\lambda$3203 \\
F475W & \ion{He}{2} $\lambda$4686, \ion{H}{1} $\lambda$4861, \ion{H}{1} $\lambda$4340,
        [\ion{Ar}{4}] $\lambda$4740, [\ion{O}{3}] $\lambda$5007, [\ion{O}{3}] $\lambda$4959 \\
F814W & Stellar Continuum, Paschen Continuum, [\ion{Cr}{5}] $\lambda$7979, \ion{He}{1}
         $\lambda$7065, [\ion{S}{3}] $\lambda$9069, [\ion{Ar}{3}] $\lambda$7135 \\
F110W & Stellar Continuum, \ion{He}{2} $\lambda$10120, \ion{He}{1} $\lambda$10830, \ion{H}{1}
        $\lambda$12820, [\ion{S}{3}] $\lambda$9532, [\ion{S}{3}] $\lambda$9069 \\
F160W & Stellar Continuum, \ion{He}{2} $\lambda$4760 \\
\enddata
\tablecomments{An extensive but incomplete list of likely flux contributors by filter.}
\label{fluxtable}
\end{deluxetable}

\footnotesize
\begin{deluxetable}{cccccccccccc}\tablewidth{0pt}
\rotate
\footnotesize
\tablecaption{M31 PNe Photometric Catalog}
\tablehead{
\colhead{{\footnotesize M06 ID}} &
\colhead{{\footnotesize PHAT RA}} &
\colhead{{\footnotesize PHAT Dec}} &
\colhead{{\footnotesize $F275W$}} &
\colhead{{\footnotesize $F336W$}} &
\colhead{{\footnotesize $F475W$}} &
\colhead{{\footnotesize $F814W$}} &
\colhead{{\footnotesize $F110W$}} &
\colhead{{\footnotesize $F160W$}} &
\colhead{{\footnotesize EC\tablenotemark{a}}} &
\colhead{{\footnotesize $z_{PN}$\tablenotemark{b}}} &
}
\startdata
26 & 11.469719 & 42.086807 & \nodata & \nodata & 24.869 & 25.920 & \nodata & \nodata & $>3.8$ & 0.35\\
29 & 11.461526 & 42.113659 & \nodata & 26.015 & 24.718 & 25.926 & \nodata & \nodata & $>3.8$ & 0.46\\
30 & 11.360975 & 42.142666 & 22.656 & 22.578 & 22.958 & 23.512 & 22.798 & 23.413 & 3.1 & 0.42\\
31 & 11.381368 & 42.144012 & 23.235 & 22.673 & 22.097 & 23.865 & 23.051 & 23.487 & $>3.8$ & 0.60\\
32 & 11.472247 & 42.147278 & 23.282 & 22.878 & 22.416 & 23.986 & 23.377 & 23.823 & $>3.8$ & 0.52\\
33 & 11.372000 & 42.149998 & \nodata & \nodata & 24.827 & 25.762 & 24.550 & 25.225 & $>3.8$ & 0.31\\
34 & 11.431614 & 42.156788 & 23.545 & 22.882 & 22.850 & 24.346 & 23.592 & 24.253 & $>3.8$ & 0.57\\
44 & 11.530962 & 42.110489 & 22.766 & 22.619 & 22.917 & 23.720 & 23.121 & 23.540 & 3.4 & 0.31\\
45 & 11.525844 & 42.116840 & 23.178 & 22.295 & 21.837 & 23.356 & 22.570 & 22.878 & $>3.8$ & 0.49\\
46 & 11.508853 & 42.127369 & \nodata & \nodata & 23.712 & 25.093 & \nodata & \nodata & 3.5 & 0.37\\
\vdots & \vdots & \vdots & \vdots & \vdots & \vdots & \vdots & \vdots & \vdots & \vdots\\
\enddata
\tablecomments{Table 2 is published in its entirety in the electronic edition of the
Astrophysical Journal. A portion is shown here for guidance regarding its form and content.} 
\tablenotetext{a}{Excitation Classification ($EC \equiv 0.45 \times (F_{5007}/F_{H\beta})$,
for $EC<5$) as estimated by $\log(F_{5007}/F_{H\beta}) = 0.7835 + 0.7003 \left( F475W - m5007
\right)$.} 
\tablenotetext{b}{The standard score of the deviation from typical PNe identified in PHAT. See Section~\ref{zPNsect} for full derivation.}
\label{pnetable}
\end{deluxetable}

\footnotesize
\begin{deluxetable}{cccc |ccccc}\tablewidth{0pt}
\footnotesize
\tablecaption{Misidentified Extended \ion{H}{2} Regions}
\tablehead{
\colhead{{\footnotesize M06 ID}} &
\colhead{{\footnotesize M06 RA}} &
\colhead{{\footnotesize M06 Dec}} &
\colhead{{\footnotesize M06 $m5007$}}\vline &
\colhead{{\footnotesize M06 ID}} &
\colhead{{\footnotesize M06 RA}} &
\colhead{{\footnotesize M06 Dec}} &
\colhead{{\footnotesize M06 $m5007$}} &
}
\startdata
35 & 11.471667 & 42.159944 & 23.90 & 52 & 11.660417 & 42.182833 & 22.93\\
53 & 11.679167 & 42.185806 & 21.31 & 54 & 11.543750 & 42.189306 & 24.33\\
55 & 11.638333 & 42.193917 & 20.60 & 56 & 11.535000 & 42.192000 & 21.03\\
57 & 11.642917 & 42.195306 & 17.75 & 58 & 11.674167 & 42.196722 & 24.10\\
59 & 11.672917 & 42.199167 & 22.39 & 60 & 11.544167 & 42.212194 & 20.97\\
62 & 11.680417 & 42.218917 & 24.26 & 64 & 11.645417 & 42.195833 & 22.08\\
84 & 11.142083 & 41.954194 & 23.41 & 96 & 11.211250 & 41.936028 & 24.84\\
103 & 11.318750 & 41.958861 & 24.43 & 106 & 11.393750 & 41.969889 & 23.70\\
109 & 11.366667 & 41.991667 & 24.30 & 117 & 11.291667 & 42.028944 & 22.73\\
120 & 11.292917 & 42.041056 & 21.21 & 121 & 11.222500 & 42.042806 & 20.83\\
125 & 11.197500 & 41.950194 & 22.46 & 126 & 11.197083 & 41.949361 & 20.92\\
139 & 11.435833 & 41.957500 & 21.99 & 147 & 11.637917 & 41.950833 & 23.49\\
150 & 11.640417 & 41.986750 & 24.02 & 163 & 11.625833 & 41.987694 & 22.78\\
241 & 11.129583 & 41.852500 & 23.86 & 242 & 11.124167 & 41.862889 & 24.08\\
243 & 11.158333 & 41.864722 & 22.23 & 246 & 11.125417 & 41.866167 & 22.65\\
257 & 11.183750 & 41.899778 & 21.29 & 262 & 11.247917 & 41.920083 & 24.35\\
268 & 11.191667 & 41.883028 & 23.50 & 279 & 11.413333 & 41.832139 & 23.39\\
285 & 11.405000 & 41.851889 & 21.10 & 295 & 11.434167 & 41.868667 & 21.20\\
298 & 11.428750 & 41.876278 & 21.67 & 301 & 11.430417 & 41.883417 & 19.67\\
307 & 11.406667 & 41.906722 & 22.84 & 322 & 11.581667 & 41.841889 & 22.11\\
324 & 11.573333 & 41.866194 & 19.19 & 327 & 11.555000 & 41.873556 & 21.66\\
446 & 11.285000 & 41.660750 & 22.30 & 448 & 11.327083 & 41.670500 & 21.64\\
481 & 11.298333 & 41.621306 & 22.08 & 482 & 11.327500 & 41.677833 & 24.12\\
494 & 11.431667 & 41.709944 & 22.32 & 659 & 11.252500 & 41.476778 & 20.03\\
660 & 11.253333 & 41.478361 & 21.43 & 661 & 11.187083 & 41.478083 & 22.74\\
739 & 11.291250 & 41.599917 & 21.51 & 741 & 11.308333 & 41.603917 & 21.42\\
954 & 10.913333 & 41.448028 & 20.35 & 996 & 10.979167 & 41.434611 & 21.93\\
999 & 11.160000 & 41.419861 & 20.89 & 1000 & 10.982917 & 41.442750 & 21.66\\
1020 & 11.197500 & 41.431500 & 21.90 & 1371 & 10.932917 & 41.193583 & 23.21\\
1377 & 10.945417 & 41.211056 & 22.46 & 1385 & 11.062917 & 41.258722 & 23.57\\
1699 & 10.895000 & 41.164944 & 19.97 & 2159 & 11.675000 & 42.262694 & 24.80\\
2581 & 11.564583 & 42.249167 & 24.53 & 2587 & 11.148333 & 41.935139 & 23.22\\
2589 & 11.416250 & 41.919278 & 23.11 & 2664 & 11.103750 & 41.627583 & 22.88\\
2669 & 11.296250 & 41.612583 & 22.18 & 2673 & 11.301667 & 41.619278 & 20.52\\
2674 & 11.299167 & 41.620250 & 20.06 & 2675 & 11.302500 & 41.621250 & 20.82\\
2678 & 11.309583 & 41.623639 & 21.73 & 2720 & 11.203333 & 41.453583 & 24.37\\
2721 & 11.187500 & 41.464028 & 20.60 & 2722 & 11.181250 & 41.465694 & 24.10\\
2994 & 11.244583 & 41.927028 & 20.15 & 3091 & 11.212917 & 41.923278 & 24.28\\
3158 & 10.895417 & 41.164583 & 19.86 & 3225 & 11.302083 & 41.621583 & 20.78\\
\enddata
\tablecomments{M06 sources that are visually extended in PHAT images and excluded from the
PHAT PNe catalog. Most are likely \ion{H}{2} regions.} 
\label{hiitable}
\end{deluxetable}

\footnotesize
\begin{deluxetable}{cccccccccc}\tablewidth{0pt}
\footnotesize
\tablecaption{Misidentified Stellar Sources}
\tablehead{
\colhead{{\footnotesize M06 ID}} &
\colhead{{\footnotesize PHAT RA}} &
\colhead{{\footnotesize PHAT Dec}} &
\colhead{{\footnotesize $F275W$}} &
\colhead{{\footnotesize $F336W$}} &
\colhead{{\footnotesize $F475W$}} &
\colhead{{\footnotesize $F814W$}} &
\colhead{{\footnotesize $F110W$}} &
\colhead{{\footnotesize $F160W$}} &
}
\startdata
65 & 11.592355 & 42.143700 & 19.167 & 19.380 & 20.748 & 20.610 & 20.683 & 20.566\\
66 & 11.620730 & 42.196083 & 19.415 & 19.446 & 20.632 & 20.228 & 20.124 & 20.020\\
100 & \nodata & \nodata & \nodata & \nodata & \nodata & \nodata & \nodata & \nodata\\
105 & 11.379783 & 41.964077 & 23.464 & 22.513 & 21.697 & 20.973 & 20.722 & 20.429\\
111 & 11.204804 & 41.991951 & 22.436 & 22.297 & 23.064 & 22.518 & \nodata & \nodata\\
116 & 11.384217 & 42.012291 & 24.817 & 22.829 & 21.228 & 19.848 & 19.489 & 19.003\\
127 & 11.369970 & 41.954872 & 20.679 & 20.608 & 21.725 & 21.205 & 21.113 & 20.974\\
151 & \nodata & \nodata & \nodata & \nodata & \nodata & \nodata & \nodata & \nodata\\
152 & 11.703743 & 42.000370 & 21.215 & 21.010 & 21.024 & 19.468 & \nodata & \nodata\\
153 & 11.630049 & 42.009483 & 22.363 & 21.906 & 22.805 & 22.122 & \nodata & \nodata\\
\vdots & \vdots & \vdots & \vdots & \vdots & \vdots & \vdots & \vdots & \vdots\\
\enddata
\tablecomments{M06 sources that are likely misidentified non-PN stellar sources and excluded
from the PHAT PNe catalog. PHAT RA, Dec, and photometry is included when available. Table 4 is
published in its entirety in the electronic edition of the Astrophysical Journal. A portion is
shown here for guidance regarding its form and content.} 
\label{sttable}
\end{deluxetable}

\end{document}